# Insight into the Correlation of Shape and Magnetism of Hematite Nanospindles

*Juri Kopp[1], Gerald Richwien[2], Markus Heidelmann[3], Soma Salamon[1], Benoît Rhein[2], Annette M. Schmidt[2], Joachim Landers*[1]*

[1] Faculty of Physics and Center for Nanointegration Duisburg-Essen (CENIDE), University of Duisburg-Essen, 47057 Duisburg, Germany

[2] Department of Chemistry & Biochemistry, University of Cologne, 50939 Cologne, Germany

[3] Interdisciplinary Center for Analytics on the Nanoscale (ICAN), University of Duisburg-Essen, 47057 Duisburg, Germany

*Corresponding author email: joachim.landers@uni-due.de*

ABSTRACT

It is established that the Morin transition, a spin reorientation in hematite, is shifted to lower temperatures with decreasing nanoparticle volume. However, our findings indicate an opposite effect in a series of hematite nanospindles: The particles, synthesized by hydrothermal decomposition of iron(III) chloride solution, with aspect ratios $p$ between 1.0 and 5.2 (long axis ca. 70 - 290 nm) display decreasing Morin transition temperatures $T_{Morin}$ upon increasing $p$, despite the volume increase. Their inner morphology, determined via (HR)STEM and XRD, shows that they are formed by the epitactical fusion of primary particles, perfectly aligned in terms of crystallographic orientation. Combining magnetometry and Mössbauer spectroscopy, we uncover the correlation between particle shape, magnetic properties, and in particular the Morin transition: While more spherical particles undergo said transition at about 200 K, $T_{Morin}$ decreases upon higher nanospindle elongation, while also being broadened and showing a wider thermal hysteresis. Our measurements reveal complete suppression of the Morin transition beyond a critical threshold $p \gtrapprox 1.5$, indicating stabilization of the weak ferromagnetic (WFM) state with net particle magnetic moment within the hematite basal plane, despite such behavior being unexpected based on shape anisotropy considerations. For the correlated, magnetic field-dependent spin-flop transition, a comparable trend in particle aspect ratio is detected. We have demonstrated the presence of intermediate spin alignment states that deviate both from the low-temperature antiferromagnetic (AFM) and high-temperature WFM spin structure for slightly elongated particles, likely being connected to the suppression of the Morin transition observed for $p \gtrapprox 1.5$.

## Introduction

Magnetic nanoparticles of all kinds, and in particular well-defined nanostructures based on iron oxides, are of steadily increasing importance in many scientific and application areas, including drug delivery,[1, 2] cancer imaging and treatment,[3, 4] magnetic resonance imaging (MRI) contrast enhancement,[5-7] catalysis, and data storage systems.[8] While the smallest nanoparticles are mostly superparamagnetic, particles with more complex magnetization dynamics and structure, such as core shell particles, Janus particles, or multicore particles etc., can have peculiar and tailorable magnetic responses.[2, 9-19] In particular, hematite ($\alpha$-$Fe_2O_3$)-based nanoparticles are easy to synthesize in various sizes and shapes.[20-22] Similar to bulk hematite, their weak ferromagnetism (WFM) observed at ambient temperatures and present up to the Néel temperature $T_N$ of ca. 948 K,[23-25] originates from a canted antiferromagnetic spin orientation within the basal plane of the rhombohedral lattice.

The latter is caused by the Dzyaloshinskii-Moriya interaction and is the origin of a small net magnetization perpendicular to the crystallographic c-axis in the hexagonal basal plane (see fig. 1 (a)).[23, 26] When cooling below the Morin transition temperature $T_{Morin}$, which for bulk hematite is typically located between 250 K and 260 K,[27-29] a spin reorientation to a purely antiferromagnetic state with spins aligned along the rhombohedral axis is observed (see fig. 1 (b)). The Morin transition temperature $T_{Morin}$ depends on different factors like crystal size, morphology, the presence and density of crystallographic defects, lattice strain, and foreign elements / dopants.[30] Most studies report a fairly constant transition temperature for hematite particles down to a diameter of about ca. 100 nm.[31] For even smaller diameters, however, $T_{Morin}$ decreases with decreasing particle size,[31-33] and for particles smaller than about 20 nm, the Morin transition is observed to disappear entirely,[31, 33] with the smallest grains staying in the WFM state down to cryogenic temperatures, which Chuev et al. also assign to effects of increasing shape anisotropy.[34] The characteristic hysteresis of the Morin transition is usually more pronounced for smaller particles, particularly in submicron-sized hematite.[31] However, depending on the particle shape, nanomorphology, and formation process, different trends in $T_{Morin}$ are described, with the underlying mechanisms not yet being completely understood.[21, 33, 35] Its investigation might result in new developments concerning devices with tailored and anisotropic physical properties with applications in nanoelectronics and spintronics.[36]

Hematite crystallizes in a rhombohedral corundum lattice, where oxygen ($O^{2-}$) anions are in a hexagonal arrangement, and iron ($Fe^{3+}$) cations occupy two-thirds of the octahedral lattice sites, [25, 37, 38] and is the thermodynamically most stable iron oxide at oxygenic ambient conditions. It is thus easily formed under various reaction conditions. In the hydrothermal treatment of aqueous iron(III) salt solutions, initially particulate iron(oxide-hydroxide) nuclei are formed. These nuclei serve in the next stage as precursor source by undergoing a peculiar Oswald-type redissolution / reprecipitation and aggregation process, that finally determines the ultimate morphology and size of the resultant hematite nanoparticles. This opens the possibility to direct the process towards preferential morphologies by proper choice and concentration of additives.

In this study, we make use of this strategy to achieve a preferential nanoparticle growth into a regular spindle shape with tailored aspect ratio *p*, as suggested by Ocana et al.[39] Their comprehensive analysis of the hydrothermal synthesis of hematite particles in the presence of sodium dihydrogen phosphate indicates that initially formed primary hematite particles subsequently aggregate into the observed spindle-shaped morphology.

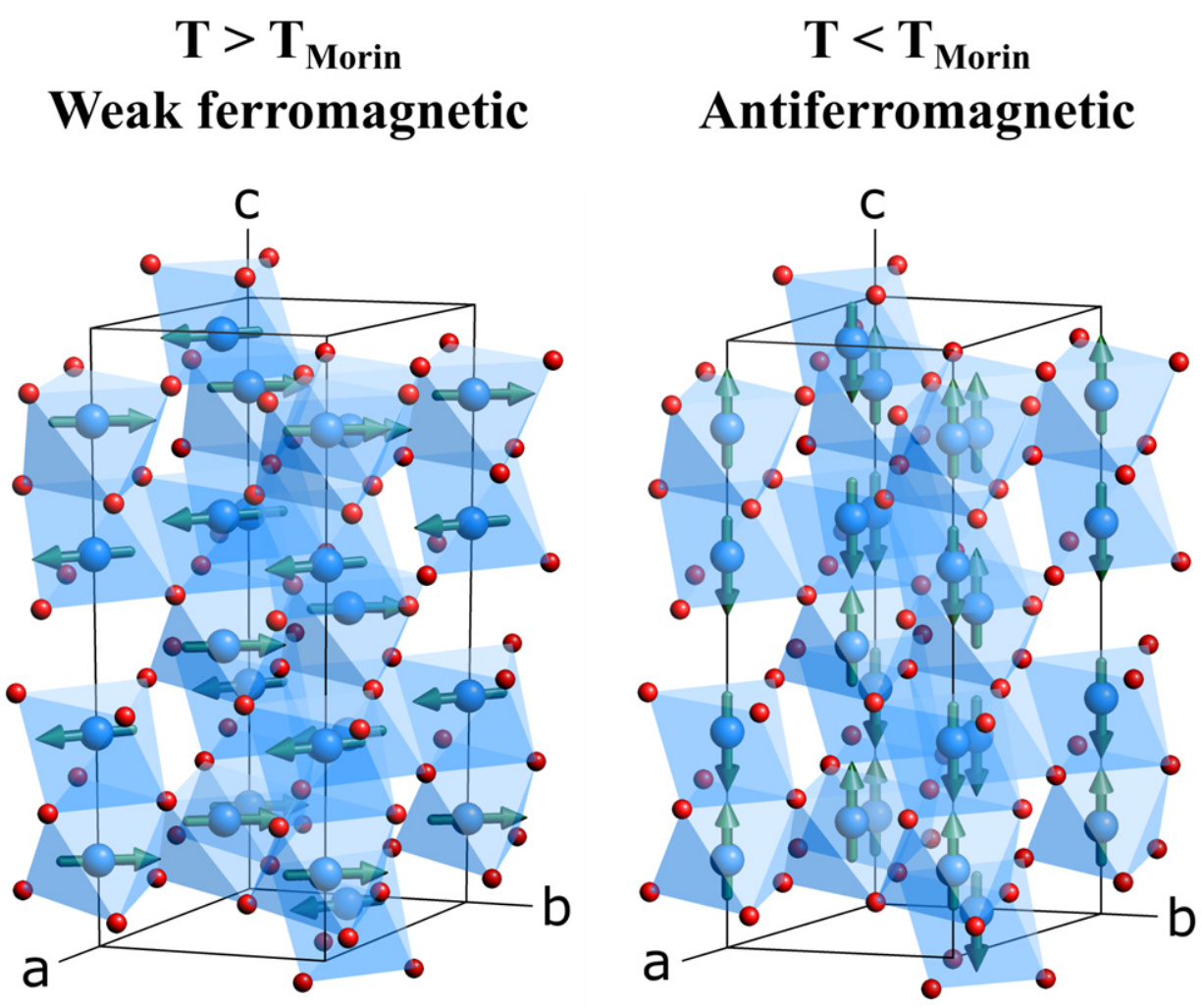

*Figure 1. Lattice structure and spin configuration of the weak ferromagnetic (WFM) phase of hematite above $T_{Morin}$ (left), and of the antiferromagnetic (AFM) phase below $T_{Morin}$ (right).*

In the following, we provide insight into the correlation between aspect ratio and magnetic properties of these spindle-type hematite nanoparticles, and show that despite increasing volume, higher particle elongation can lead to a complete suppression of the Morin transition in these structures. To provide a comprehensive understanding, we present results on a series of hematite nanospindles of varying aspect ratio, prepared by the hydrothermal method. Using a combination of methodologies including (high resolution) scanning transmission electron microscopy ((HR)STEM), x-ray diffractometry (XRD), (thermo)magnetometry and Mössbauer spectroscopy, we provide a closer insight to the physical background and origin of this unexpected behavior of the Morin transition.

## Experimental methods

**Chemicals:** Iron(III)chloride hexahydrate ($FeCl_3 \cdot 6H_2O$, 99 %) and tetramethylammonium hydroxide (TMAH, 25 % in water), are purchased from Sigma Aldrich (Darmstadt, Germany). Sodium dihydrogenphosphate monohydrate ($NaH_2PO_4 \cdot H_2O$, 98 %) is bought from Alfa Aesar (Kandel, Germany). Iron(III)nitrate nonahydrate ($Fe(NO_3)_3 \cdot 9H_2O$, laboratory reagent grade) and nitric acid ($HNO_3$, 65 wt-% in water, technical grade) are purchased from Fisher Scientific (Schwerte, Germany). Citric acid monohydrate (≥ 99.5 %) is bought from Jungbunzlauer GmbH (Ladenburg, Germany).

**Synthesis:** Spindle-like hematite ($\alpha$-$Fe_2O_3$) particles are obtained via the hydrothermal method based on the synthetic route of Ozaki et al.[40] 1 L of distilled water is heated to 100 °C in a three-necked flask with a reflux condenser. 20 mmol $l^{-1}$ $FeCl_3 \cdot 6\,H_2O$ and 0.6 mmol $l^{-1}$ ($p = 5.2$), 0.3 mmol $l^{-1}$ ($p = 3.8$), 0.2 mmol $l^{-1}$ ($p = 2.5$), 0.075 mmol $l^{-1}$ ($p = 1.5$), 0.05 mmol $l^{-1}$ ($p = 1.25$) and 0.0 mmol $l^{-1}$ ($p = 1.0$) $NaH_2PO_4 \cdot H_2O$ is added to the reaction mixture. The solution is stirred under reflux for 48 h. After cooling to room temperature, the particles are collected via centrifugation (8500 rpm, 15 min) and washed with water five times and finally dispersed in water. The particles are electrostatically stabilized with citric acid following the protocol of Wagner et al. [9] To a dispersion of $\alpha$-$Fe_2O_3$ in 50 mL water (ca. 3.5 wt%), 50 mL of 2 M $HNO_3$ is added and stirred 5 min at room temperature. Afterwards, 50 mL of an aqueous solution of $Fe(NO_3)_3$ (0.35 M) is added and the reaction is stirred to reflux for 1 h. After cooling to room temperature, the particles are collected via centrifugation (6500 rpm, 15 min) and washed one time with 2 M $HNO_3$ and three times with water. To the resulting dispersion of $\alpha$-$Fe_2O_3$ in water, 0.1 M citric acid solution is added dropwise until the particles precipitate, followed by the addition of tetramethylammonium hydroxide to set the pH to 8 – 9. Finally, the particles are washed two times with water and dispersed in water.

**Transmission electron microscopy:** Transmission electron microscopy images were recorded by a LEO 912 from the company Zeiss with a $LaB_6$ cathode working with a voltage of 120 kV. A highly diluted solution of the sample was prepared and applied to a Carbon-based Quantifoil Multi A TEM grid and the solvent was slowly evaporated. The images were then recorded by a Moorweis Slow Scan CCD camera Sharp:EYE 2048 x 2048 TRS. The long and short axes were manually measured for over 300 particles to determine the size and aspect ratio by calculating the ratio of long axis to short axis. High-resolution STEM measurements were carried out utilizing a probe-side Cs-corrected JEOL JEM 2200fs operated at 200 kV acceleration voltage.

**X-Ray Diffractometry:** For X-ray diffraction measurements performed at room temperature, we use a Philips PW1730 commercial θ-2θ X-ray diffractometer. It is equipped with a horizontal goniometer, a Cu anode X-ray tube, and a graphite monochromator.

**Magnetometry:** Field- and temperature dependent magnetization measurements of hematite nanospindle powders are performed utilizing the vibrating sample magnetometer (VSM) option of a Quantum Design PPMS DynaCool. M(H) curves are recorded between 4.3 K and 300 K up to maximum fields of ±9 T, zero-field-cooled/field-cooled (ZFC-FC) magnetization curves are recorded in the same temperature interval at 10 mT.

**Mössbauer Spectroscopy:** A Spectromag 4000-10 magnet cryostat from Oxford Instruments is used for magnetic field-dependent Mössbauer spectroscopy at ca. 4.3 K up to 10 T, with the sample located in a split-pair system of superconducting coils. Mössbauer spectra without applied magnetic field are recorded between 5 and 300 K using a closed-cycle cryostat setup (SHI-850-5) by Lake Shore Cryotronics. For the measurements we utilize a $^{57}$Co(Rh) source installed in a Mössbauer velocity transducer operated in constant acceleration mode. To prevent texture effects due to the potential alignment of elongated nanospindles when forming thin powder films, the hematite particles are pressed to pellets containing ca. 20 mg/cm$^2$ of hematite nanospindles mixed with ca. 200 mg of chemically inert boron nitride. For the temperature-dependent series of spectra, data are reproduced using a multi-level relaxation approach to consider effects of spectral deformation by beginning superparamagnetic dynamics.[41] For spectral components of less than ca. 10% total spectral area, hyperfine parameters are fixed to extrapolated values to minimize the error due to overfitting the data. To account for the moderate sample thickness, which influences the spectral shape, we adopt the method outlined in Ref.[42]

## Results and Discussion

### Particle formation process & morphology

Hematite particles in various sizes and shapes are available via different synthetic routes.[20, 22, 43] Among these, spindle-shaped nanoparticles attracted much interest due to their unique perpendicular configuration between geometric axis and the preferential macrospin orientation. Here, we investigate the interplay between geometry, magnetic anisotropy, and thermomagnetic properties for spindle-shaped hematite nanoparticles synthesized by a hydrothermal process through the hydrolysis of iron(II) chloride aqueous solution in the presence of sodium dihydrogen phosphate.[40] By varying the concentration of the latter (see experimental section), it is possible to tailor the aspect ratio $p$ of the resulting particles between 1 and 6 (or higher): The phosphate anions adsorb to the surface of the nanoparticles, thereby preferentially occupying surface sites parallel to the rhombohedral c-axes. As a consequence, this site-selective surface complex formation leads to a preferred growth of the particles along the rhombohedral axis by aggregation of the primary particles. This unique growth process leads to epitactically aggregated structures with a spindle shape and tunable aspect ratio.[39] For the current investigation, six batches of spindle-shaped hematite nanoparticles with aspect ratios between 1.0 and 5.2 are synthesized by varying the phosphate concentration accordingly (see fig. S1). We observe a direct and linear correlation of aspect ratio p on $NaH_2PO_4$ concentration, ensuring high sample reproducibility.

After the thermolysis process, the particles are collected and washed carefully, and particle shape and morphology are analyzed in detail by TEM. At 10,000x magnification (fig. 2) the general morphology, uniformity, size, and aspect ratio of the particles can be evaluated, with main results being summarized in Tab. 1. Histograms on the length distribution of both geometric axes indicate a moderate distribution of both values around 15 %. For all samples, the short particle axis length $x$ is relatively similar around 70 nm. At the same time, the long axis length $y$ increases from ca. 70 nm for the nearly spherical batch synthesized in phosphate-free solution to 290 nm for particles formed at the highest phosphate concentration, thus confirming the validity of the approach, and that the particle volume can be considered as approximately proportional to the aspect ratio.

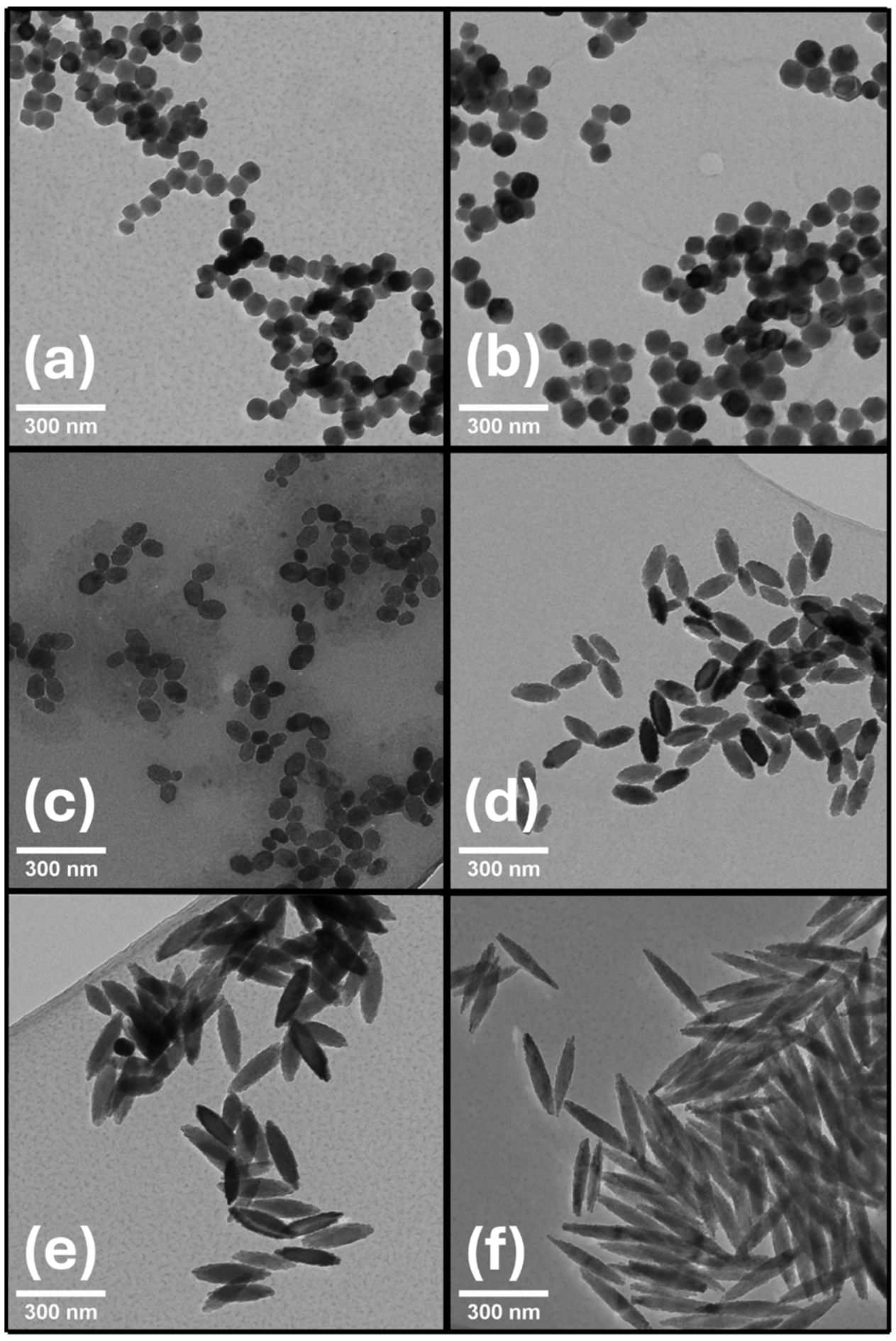

*Figure 2. TEM images of hematite nanospindles with an aspect ratio p between 1.0 (a) and 5.2 (f). Recorded at 10000x magnification, scale bar 300 nm.*

*Table 1. Long axis length y, short axis length x, and aspect ratio p of hematite nanospindles as determined from TEM data. Standard deviation given in brackets.*

| **phosphate conc. *c* (mmol/l)** | **long axis length *y* (nm)** | **Short axis length *x* (nm)** | **aspect ratio *p*** |
|---|---|---|---|
| 0.00 | 68.6(4.2) | 68.6(4.2) | 1.0 (0) |
| 0.05 | 107.1(2.6) | 85.3(2.5) | 1.25 (0.14) |
| 0.075 | 84.5(7.3) | 58.6(5.4) | 1.5 (0.12) |
| 0.20 | 144.9(7.2) | 57.9(3.8) | 2.5 (0.13) |
| 0.30 | 229.4(10.7) | 60.3(3.8) | 3.8 (0.19) |
| 0.50 | 292.0(12.9) | 56.4(4.0) | 5.2 (0.25) |

In the detailed analysis of the particle morphologies based on TEM data, several key observations emerge: First of all, for the sample synthesized under phosphate-free conditions with an aspect ratio of 1.0, the particle shape deviates from a perfect sphere towards a rounded cuboidal shape. Especially for particles of larger aspect ratio, upon a closer inspection of the surface structure and the internal morphology of individual nanoparticles displayed in Fig. S2, the smaller, slightly elongated primary particles are still discernable by the shape of the particle edges. The interplay and interactions of these primary subunits in the resultant nanospindles in regard to crystallographic and magnetic properties are expected to strongly affect the magnetic behaviour of the individual nanospindle as well as their ensembles, and therefore have to be considered in the further evaluation.

**Crystal structure analysis**

In order to allow a more in-depth analysis of the particle nanostructure and crystallinity, we employ XRD, and (HR)STEM. Fig. 3 shows the XRD diffractograms measured for hematite nanospindles ($p$ = 1.0 to $p$ = 5.2). All observed Bragg reflexes can be assigned to the rhombohedrally centered hexagonal hematite structure ($R\bar{3}c$ space group), consisting of hexagonally closely packed oxygen layers, with two-thirds of the octahedral interstices occupied by $Fe^{3+}$ ions. No signs of additional, parasitic phases are visible and diffractograms of all samples are similar in structure. Still, minor differences are visible upon rising aspect ratio, showing a continuous change in relative intensities, which we assign to texture effects, as elongated hematite nanospindles assume an increasingly horizontal orientation on the substrate, used here for XRD measurements. Based on reflex positions, we find average lattice parameters of $a \approx 5.03$ Å and $c \approx 13.77$ Å, similar to literature values,[44] with the latter showing a slight increase for higher aspect ratios, which could be connected to a higher number of defects lying in the [001] direction, i.e. the long particle axis, as discussed by Ocana et al.[39] Analyzing the width of the (204) reflex using the Scherrer equation, we find crystallite sizes of ca. 27 nm, without any trends for different nanospindle aspect ratios. This matches previous reports on nanospindles prepared via the same synthesis routine,[39] where crystallite size was associated with the diameter of primary particles during growth of the nanospindles. Furthermore, when comparing the width of different Bragg reflexes, no increase towards higher diffraction angles is found. The latter would indicate considerable microstrain in the particles, known to affect magnetic anisotropy, whereby its absence is of relevance for the later on discussed Morin transition.[45]

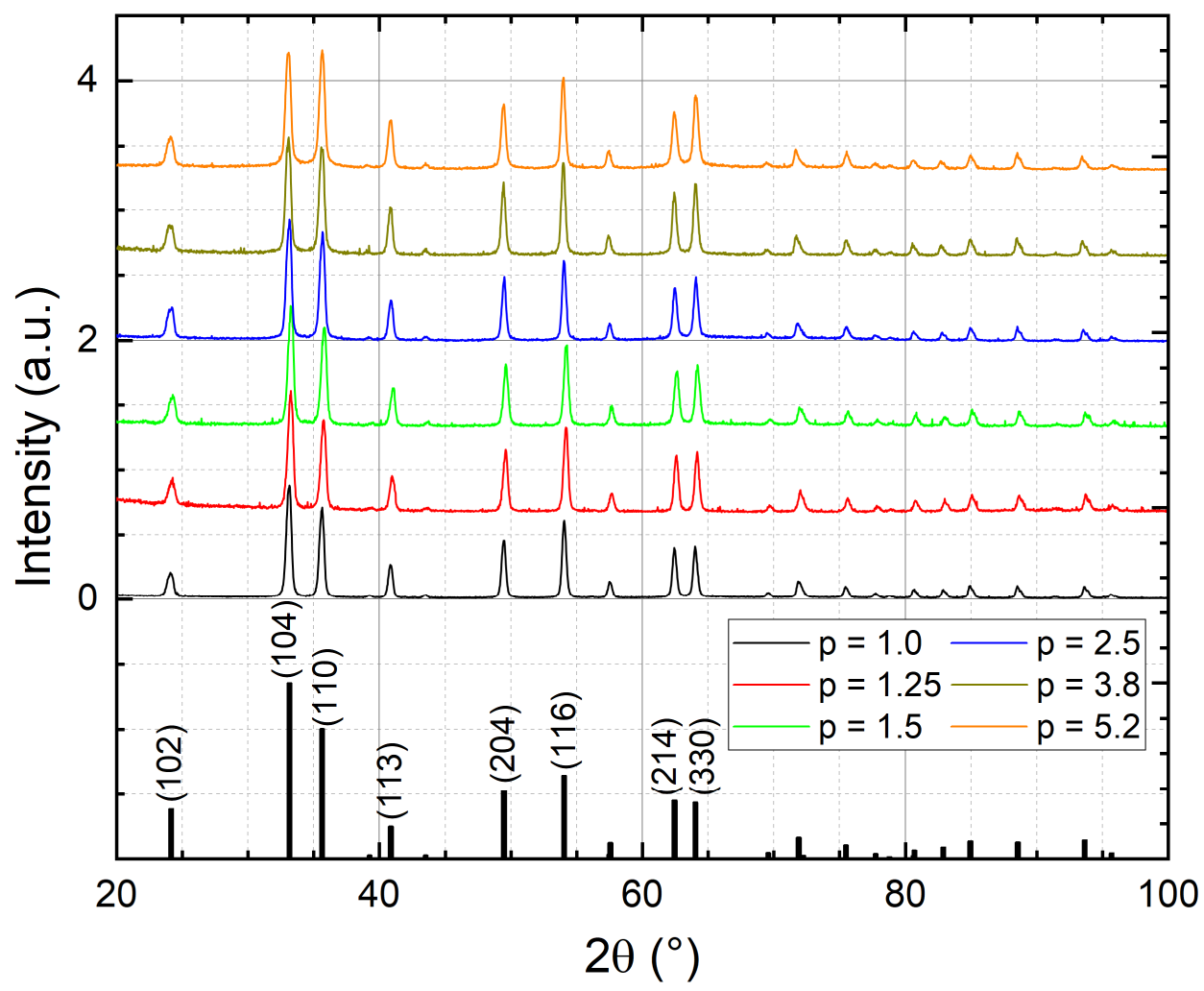

*Figure 3. XRD data recorded for hematite nanospindles with aspect ratio p = 1.0 to 5.2. Bar diagram: hematite reference data (ICSD 15840 [44]).*

In addition, images by high-angle annular dark-field (HAADF) STEM of representative nanoparticles are shown in fig. 4, with atomic columns appearing as white spheres on dark background in this imaging mode. Direct comparison with the crystal structure of hematite (inlays in the bottom right)[44] does not reveal any discrepancies. The most striking observation is that although the spindles consist of smaller primary particles, the epitactic order of the crystal lattice does not only extend to the field of view, but throughout the complete hematite nanospindle: Several random positions are imaged in high-resolution, to ensure that the same crystal orientation is present throughout the spindle. Fourier transformed images of fig. 4 confirming the epitaxial nature of the nanospindles can be found in fig. S3. Moreover, (HR)STEM images exhibit regions of slightly darker contrast, indicating less material within the electrons' path. These can be seen most prominently in fig. 4 c and d, there displaying irregular shape and varying dimensions of some nm. These observations indicate the presence of small pores, most likely being remnants from the particle assembly process. Similar observations have been reported by Ocana et al.[39]

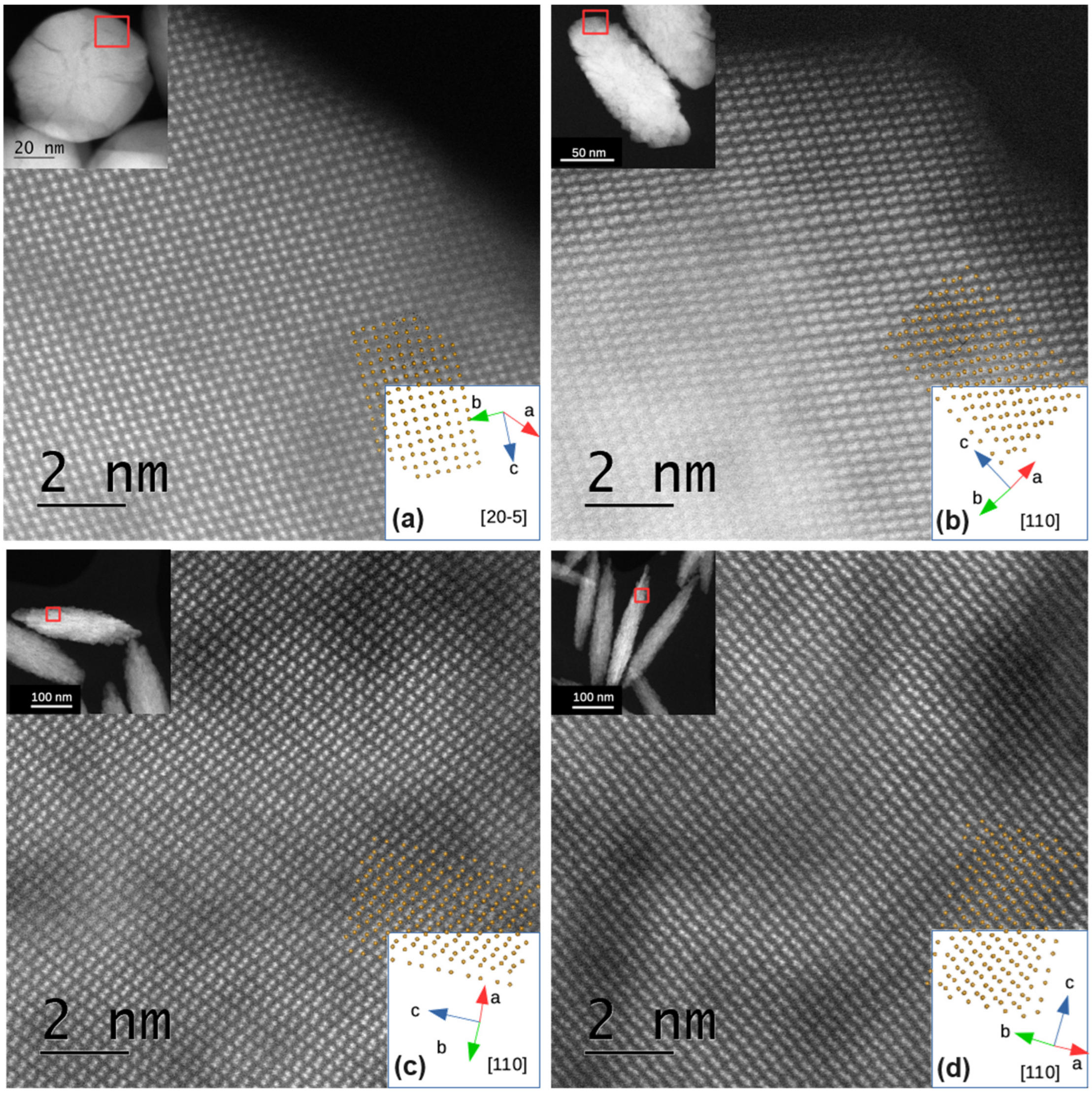

*Figure 4. High-resolution STEM images of spindle-shaped nanoparticles with aspect ratio p =1.0 (a), p = 2.5 (b), p = 3.8 (c) and p = 5.2 (d). Magnified regions are marked as red squares in the insets on the upper left corner, lower right insets show the individual particles' crystallographic orientation.*

**Magnetic characterization**

A first impression of the effect of elongated shape on the magnetic structure of the hematite nanospindles is obtained from field-dependent magnetization curves recorded at 300 K (fig. 5) and 4.3 K (fig. 6). At 300 K, we observe *M*(*H*) loops typical for the hematite state above the Morin transition, showing only minor variations in WFM saturation magnetization of ca. 0.33 $Am^2/kg$ and magnetic susceptibility, as summarized in table S1. Focusing on the shape of magnetic hysteresis (s. fig. 5b), we observe decreasing trends in coercivity $H_C$ and remanence $M_R$ for decreasing aspect ratio p. These can likely be attributed to stronger effects of beginning superparamagnetic relaxation in nanospindles of lower volume, i.e. aspect ratio *p*, minimizing $H_C$ and $M_R$. However, a comparison to the *M*(*H*) data in fig. 6 recorded at 4.3 K is necessary to evaluate whether the particles at 300 K only exhibit different degrees of superparamagnetism, or whether intrinsic differences in magnetic behavior are also present:

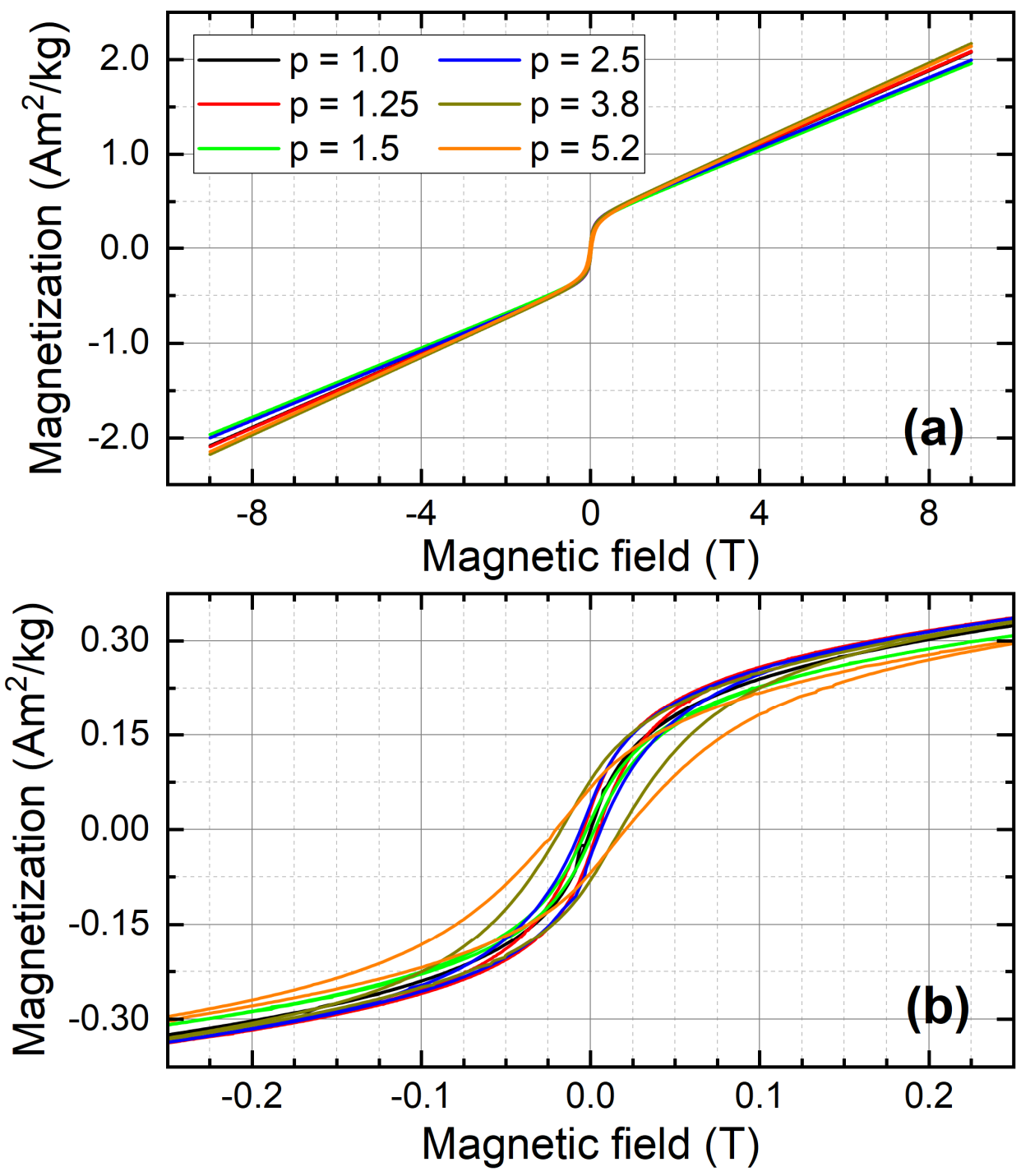

*Figure 5. M(H) curves for powder samples of nanoparticles with aspect ratio p = 1.0 to 5.2 (a) recorded at 300 K up to maximum magnetic fields of ±9 T, shown in (b) in magnification.*

At 4.3 K, the general behaviour of the samples can be divided into two groups: While particles with a high aspect ratio ($p > 1.5$), show a single s-shaped and hysteretic loop, samples with a lower aspect ratio show a more complex behaviour, assigned to the presence of a field-induced magnetic transition. In more detail, particles of low aspect ratio ($p = 1.0$ to 1.5) display the antiferromagnetic state in the low-field region, while at spin-flop transition fields $B_{SF}$ of ca. 4.8 T a field-induced transition to the WFM-state is observed. As illustrated, e.g., by Bakkaloglu et al., hematite performs a spin-flop transition for fields $B$ parallel to the c-axis at $B_{SF}$ of ca. 6.5 T, while for perpendicular fields a continuous screw-rotation reorientation is found at ca. 16 T, resulting in high-field WFM states of different spin orientation.[46-48] Zysler et al. found a distinct decrease in $T_{Morin}$ and $B_{SF}$ for lower crystallite sizes in hematite nanoparticles,[49] however, as illustrated above in XRD and TEM analysis, no relevant trend in crystallite size was found for our particles of different aspect ratio. While for aspect ratio $p = 1.0$ the transition region is relatively narrow and shows a minor hysteresis loop of ca. 0.3 T width, it is continuously broadened and shifted to lower fields upon rising particle elongation. For $p \geq 2.5$, on the other hand, particles already exhibit the WFM state without external magnetic field and show only minor variation in the shape of magnetic hysteresis, verifying the suppression of the field-induced transition for particles with an aspect ratio between 2.5 and 5.2, as shown in fig. 6 (b). A similar observation was reported by Wang et al. on the suppression of the Morin transition in hematite nanotubes as compared to nanorings.[35] In dependence of the aspect ratio of the particles, the remanent magnetization $M_R$ slightly decreases, while the coercivity $H_C$ shows a minor increase upon rising aspect ratios. It is interesting to note that particles with aspect ratio 1.5 do not completely re-attain the AFM state even after returning to zero magnetic field. A detailed investigation of the field-induced AFM-WFM transition upon rising temperature is shown in fig. S4 of the Supplemental information for particles of aspect ratio $p$ = 1.0, displaying a trend of $B_{SF}(T)$ decreasing slowly at low temperature and increasingly fast when approaching $T_{Morin}$, consistent with experiments reported by Shapira et al. on hematite single crystals.[46]

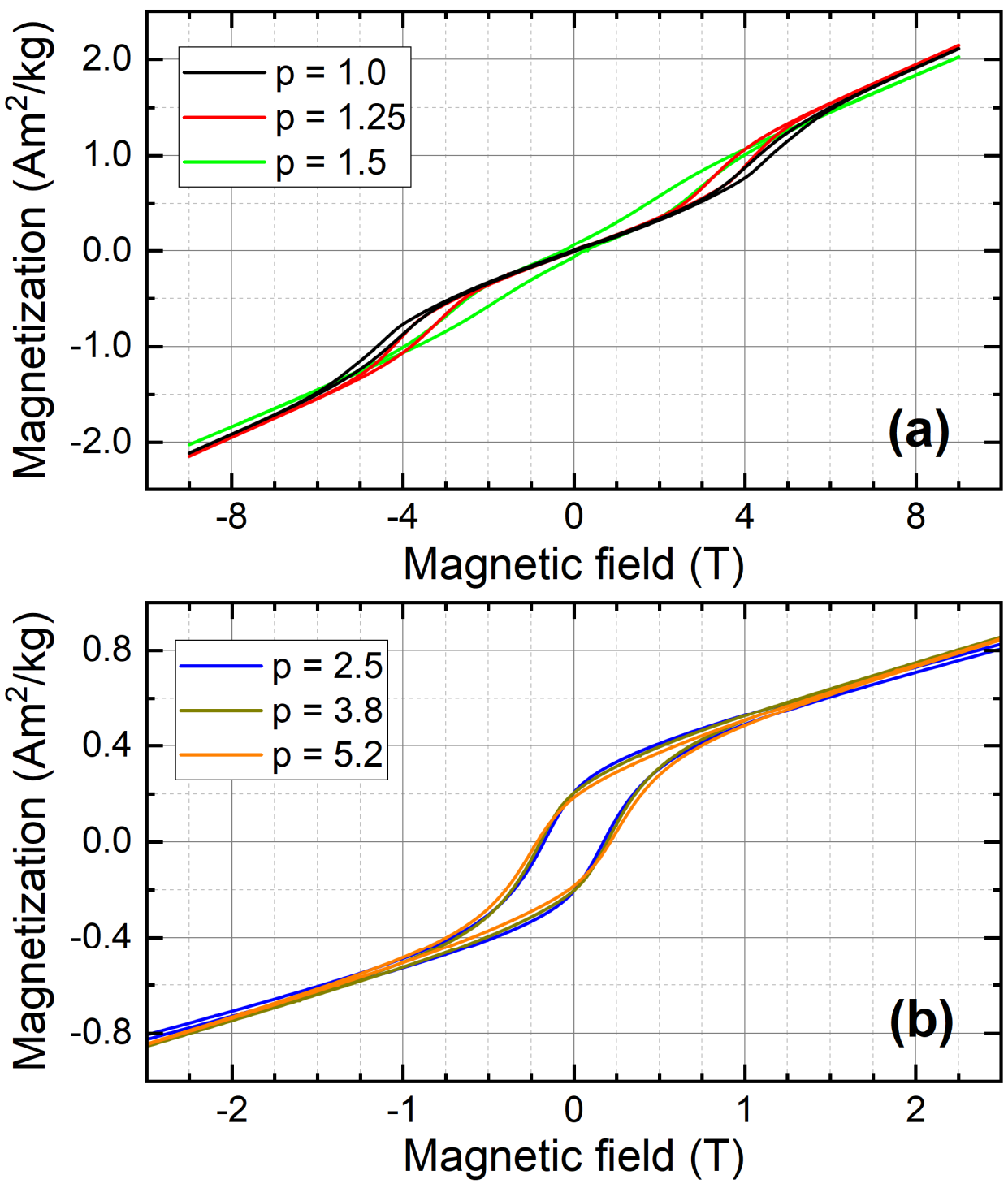

*Figure 6: M(H) curves for powder samples of nanoparticles with aspect ratio p = 1.0 to 1.5 (a) and 2.5 to 5.2 (b) measured at ca. 4.3 K at maximum magnetic fields of ± 9 T.*

To evaluate the influence of the nanospindle structure and elongation on the Morin transition as well as the particles' superparamagnetic behavior, zero-field-cooled/field-cooled (ZFC/FC) magnetization curves were recorded at 10 mT as displayed in fig. 7. For aspect ratio 1.0 (fig. 7 (a)), we observe a relatively narrow Morin transition at $T_{Morin}$ ~ 213 K upon heating, which shows a ca.

17 K wide thermal hysteresis. This temperature region matches previous reports on nanoparticles of 50 nm – 100 nm,[31] with $T_{Morin}$ already being strongly reduced relative to the bulk value of 263 K due to size effects. Above $T_{Morin}$, the particles in the WFM state partially display superparamagnetic relaxation, resulting in a splitting of the ZFC- and FC-branches of the magnetization up to the maximum temperature of 300 K. For particles with an aspect ratio of $p$ = 1.25, $T_{Morin}$ is further decreased to ca. 207 K and the thermal hysteresis broadened to a width of 27 K. With an increase in aspect ratio to $p$ = 1.5, a transitionary behavior is observed instead: $T_{Morin}$ is lowered to about 168 K, while the splitting of the ZFC and FC branches in the superparamagnetic/WFM region is more pronounced, but more importantly, we observe a considerable and almost constant FC magnetization in the low temperature range down to 5 K. A possible origin for this low-temperature FC magnetization is discussed in the context of Mössbauer spectroscopy results obtained for this sample.

As evident in fig. 7(b), the thermally induced Morin transition, similar to the corresponding field-induced process, is suppressed for higher aspect ratios starting around $p$ = 2.5. Instead, we observe a trend of decreasing splitting of the ZFC/FC branches of the magnetization for samples with higher aspect ratio, probably in connection to the higher particle volume of the nanospindles in these samples and to an influence of the elongated particle shape on the effective magnetic anisotropy. This observation is in correspondence with the marked decrease in the remanent magnetization and coercivity for particles with lower $p$ in field-dependent magnetization measurements at 300 K, pointing towards beginning superparamagnetic relaxation.

Discussing magnetic relaxation in more detail, one has to keep in mind some aspects of the hematite magnetic structure as well as the nanostructure of the studied spindle particles: Early FMR studies on hematite single-crystals already revealed almost negligible anisotropy energy barriers between the triaxial easy directions in the hematite basal plane,[50, 51] which in general would enable beginning superparamagnetism even for particles as large as the used nanospindles. This type of superparamagnetic relaxation limited to reorientation in the basal plane was discussed e.g. by Kündig et al. for hematite particles below 18 nm, reporting effective anisotropy energy constants of ca. 4 $kJ/m^3$.[52] Based on this number, on the other hand, superparamagnetic relaxation would have to be assigned to smaller superspins connected e.g. to the former primary particles instead of those of the larger hematite nanospindles. Thereby the question has to be raised, whether the nanospindles' magnetic characteristics have to be understood by collective magnetic behavior of the whole particle or rather by an assembly of strongly interacting magnetic subunits, as described by Reufer et al. for considerably more porous nanospindles.[53] Considering the limited number of larger pores in the nanospindles at hand, based on TEM-images, suggesting rather extensive contact areas between the regions corresponding to individual primary particles previous to recrystallization, it seems more natural to visualize a collective magnetic state. However, this question will also be addressed in the following section under regards of field-dependent Mössbauer spectroscopy experiments.

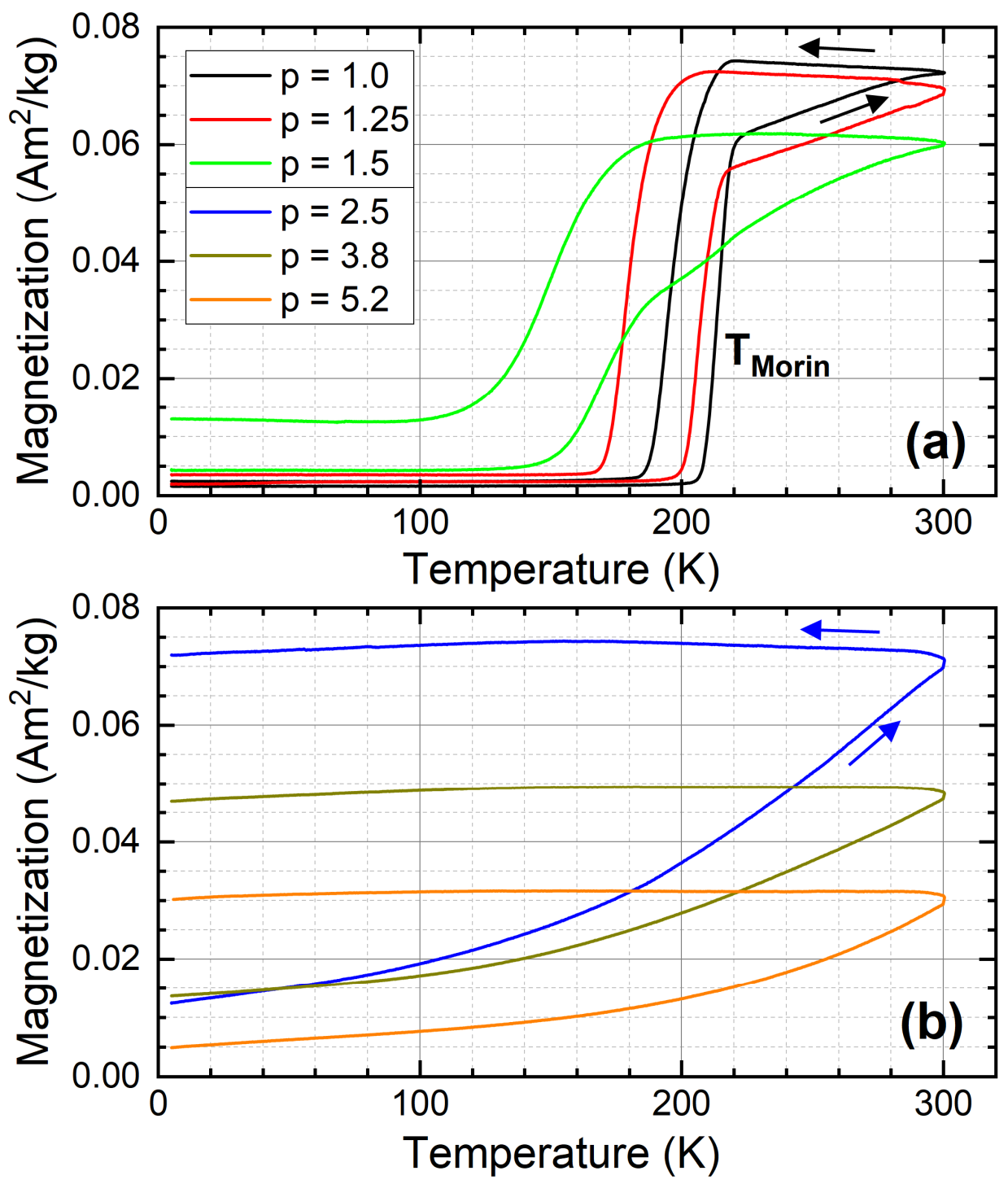

*Figure 7. ZFC/FC magnetization curves recorded at 10 mT for aspect ratio p = 1.0 - 1.5 (a) and 2.5 - 5.2 (b).*

## Mössbauer spectroscopy

While magnetometry experiments provide valuable insight into presence, shape and width of the Morin transition and its thermal hysteresis, the superposition of $T_{Morin}$ with the onset of superparamagnetic nanospindle relaxation hinders a more thorough analysis of both effects. For that purpose, we utilize Mössbauer spectroscopy at temperatures of 4.3 to 300 K and in magnetic fields up to 10 T, directly correspondent to the temperature- and field-dependent magnetometry measurements shown above.

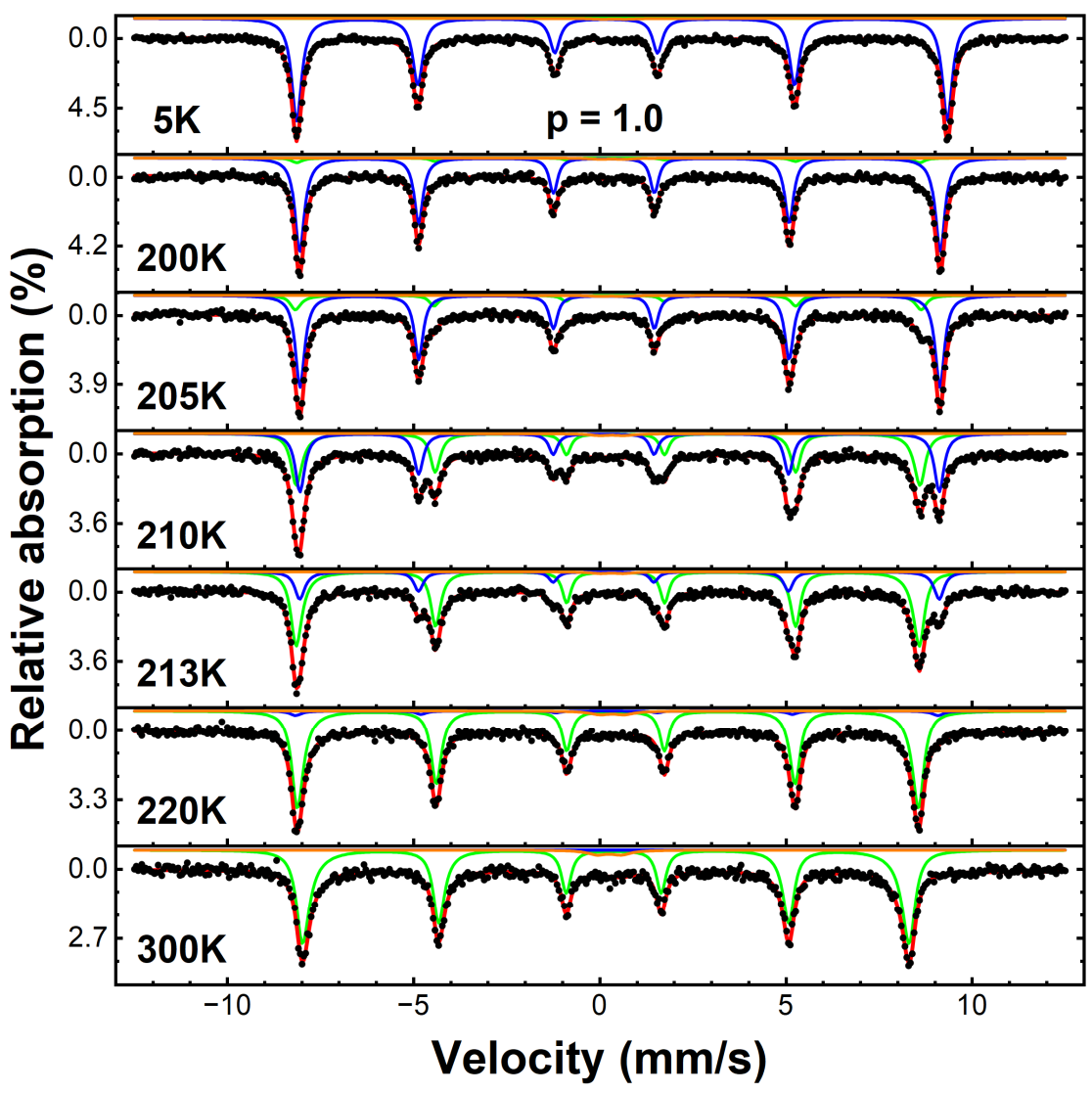

*Figure 8. Mössbauer spectra of sample p = 1.0 recorded from 5 K to 300 K across the Morin transition at ca. 210 K. Experimental spectra are reproduced using a multilevel relaxation approach to model effects of beginning superparamagnetic dynamics.*[41]

Spectra recorded between 5 K and 300 K are shown in fig. 8 for $p$ = 1.0 and reproduced with two asymmetric sextets using a multilevel relaxation approach, as described in Ref. [41], representing the low-temperature AFM (blue) and high-temperature WFM (green) state of hematite. The relatively narrow Morin transition centered at ca. 210 K is clearly visible, in agreement with ZFC-FC magnetometry experiments on this sample, and displays WFM and AFM states in coexistence in the region of thermal hysteresis.[31, 54]

As the temperature approaches 300 K, an additional very minor doublet component (orange) becomes apparent, presumably representing fast superparamagnetic relaxation. The doublet spectral area never exceeds ca. 3 %, agreeing with the observation of only slightly deformed sextet absorption lines, both indicating beginning rather than fast superspin dynamics up to ambient temperature in most of the hematite nanospindles. Spectra seem to indicate a slightly more pronounced deformation for spindles of aspect ratio of 5.2 (compare fig. 8 and S5), which counterintuitively would imply faster relaxation of larger particles. However, these differences are very minor and the studied temperature region far below $T_B$, being higher in Mössbauer spectroscopy due to its shorter characteristic timescale, so one has to ask whether a direct comparison to information on superparamagnetic relaxation inferred from ZFC-FC magnetization data shown in fig. 7 is possible. Mørup et al. discussed first changes in spectral structure of nanoparticles upon rising temperature in terms of a thermally excited oscillation of the particle magnetic moment within the energy minimum of an easy magnetic direction rather than superspin flips,[55] unlike superparamagnetic relaxation in ZFC-FC magnetometry. As furthermore the maximum achievable temperatures of both methods that can be attained without permanent changes to the particles' nanostructure are far below the average blocking temperature $T_B$, we have to abstain from a complete analysis of relaxation dynamics of the hematite nanospindles, which would allow further interpretation on the interplay of particle shape and effective magnetic anisotropy from this perspective.

Corresponding series of Mössbauer spectra are gathered for $p = 1.5$ and $p = 5.2$ in fig. S5 of the supplementary material, representative for nanospindles of moderate and high aspect ratio, respectively. To illustrate the change in Morin transition behavior more clearly, spectral areas of the WFM state are additionally compared in fig. 9 (a) for all samples. In agreement with ZFC/FC magnetometry, $T_{Morin}$ shifts gradually to lower temperatures for moderate aspect ratios up to 1.5, accompanied by a widening of thermal hysteresis. Also, we observe an increasing deformation of the previously relatively symmetric thermal hysteresis for $p = 1.5$, which could not be demonstrated clearly in magnetometry due to the overlap with ZFC/FC splitting at the onset of superparamagnetic relaxation. Of special interest here is the WFM spectral component, remaining at ca. 20 % from 100 K down to the minimum temperature, verifying that the sample does not attain the AFM ground state even far below $T_{Morin}$. For any higher aspect ratio (2.5 - 5.2), no AFM sextet is visible at any temperature, indicating the complete suppression of the Morin transition, confirming the magnetometry findings.

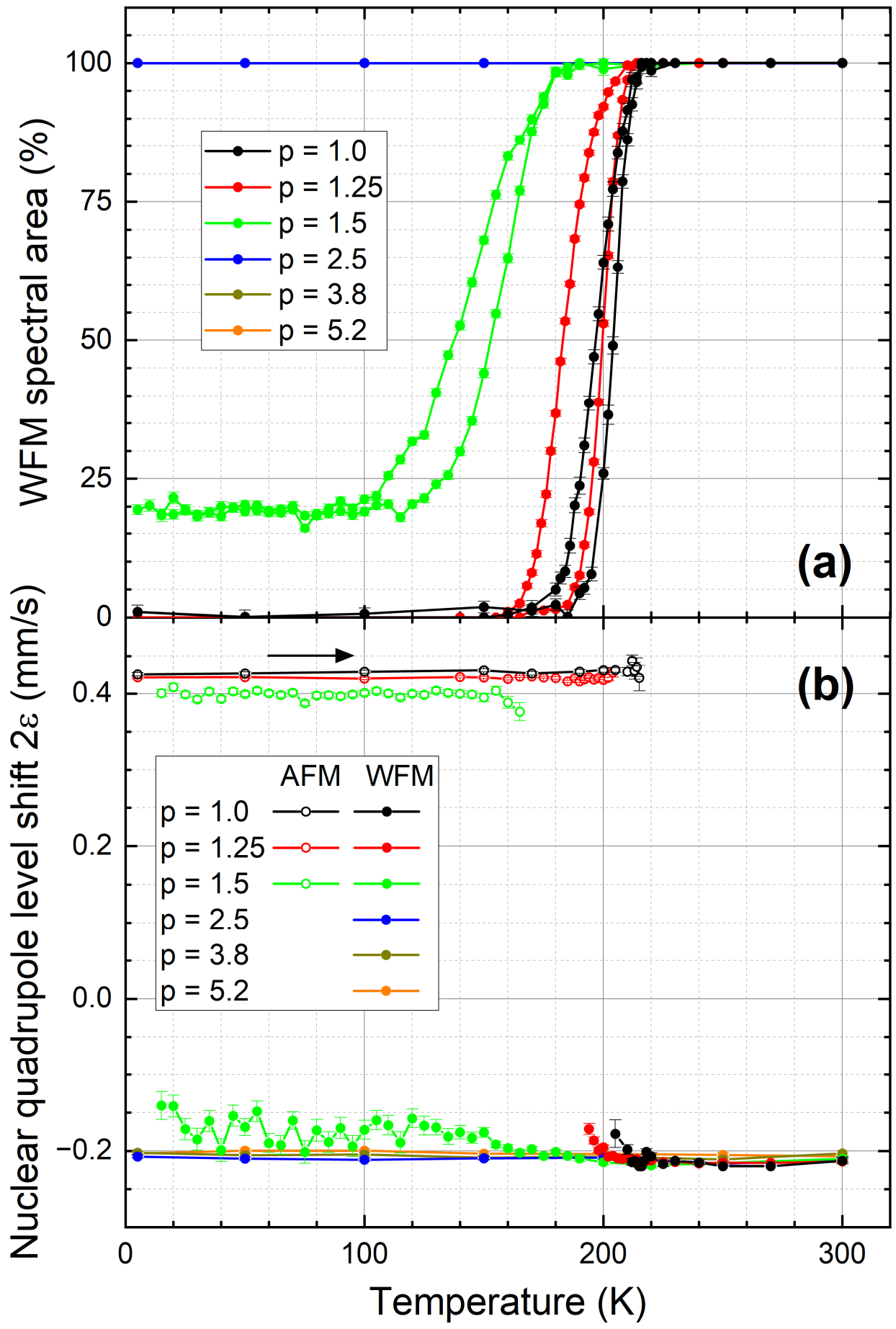

*Figure 9. (a) Spectral area of the high-temperature WFM contribution determined from Mössbauer spectra in fig. 8, showing the decrease in Morin transition temperature for p = 1.0 - 1.5, followed by a complete suppression for higher particle elongation. Spectral contributions are normalized to the total area of magnetically blocked sextet area, the spectral area of the superparamagnetic doublet, which cannot be clearly assigned, is not considered. (b) Nuclear*

*quadrupole level shift 2ε for p = 1.0 - 5.2 in the WFM (full symbols) and AFM state (empty symbols) upon rising temperature.*

As the nanospindles show such a continuous trend in suppression of the Morin transition upon particle aspect ratio, it stands to reason that the WFM magnetic moments conserved for $p$ = 1.5 down to 5 K are part of nanospindles exceeding the average aspect ratio of this sample. This would point towards a threshold value in the particle shape, above which the Morin transition no longer takes place.

Studying the hyperfine parameters extracted from data in fig. 8 in more detail, we find close resemblance to the peculiar behavior for $p$ = 1.5: The nuclear quadrupole level shift 2ε, displayed in fig. 9 (b) allows us to determine the precise angle between the spins and the electric field gradient. We yield average 2ε values of ca. 0.42 mm/s for particles with aspect ratios of $p$ = 1.0 and 1.25 in the AFM state and -0.21 mm/s in the WFM state, where $2\varepsilon_{AFM}/2\varepsilon_{WFM}$ = -2 verifies a spontaneous spin-flop transition by 90°, with $2\varepsilon_{AFM}$ being the $2\varepsilon$ value in the AFM state and $2\varepsilon_{WFM}$ in the WFM state respectively.[48] Accompanying the Morin transition is a spontaneous decrease by ≈ 0.90 T in the hyperfine magnetic field $B_{HF}$(T), when spins change alignment from 0° to 90° relative to the hexagonal c-axis. This is consistent with findings by Ruskov et al, matching this change in the anisotropic hyperfine field to the combined orbital and dipolar contributions,[27] while no considerable change in isomer shift $\delta$(T) is found (see fig. S6). For particles with aspect ratios between $p$ = 2.5 and 5.2, $2\varepsilon$ is constant within the error margin in the entire studied temperature range. $p$ = 1.5, on the other hand, with ca. 80 % of spins partaking in the Morin transition, shows a slightly lower $2\varepsilon_{AFM}$ of ca. 0.40 mm/s, whereby $|\frac{\varepsilon_{AFM}}{\varepsilon_{WFM}}| < 2$ could indicate a change in spin direction by less than 90°. The remaining 20 % of spectral area do not display the 90° spin-flop,

but instead attain a slightly different average value $2\varepsilon_{WFM} = -0.17$ mm/s below $T_{Morin}$, showing a simultaneous but minor out-of-plane alignment of spins, similar to intermediate states of spin alignment during field-induced AFM-WFM transition as noted in previous studies.[47]

In an attempt to gain further information on the origin of the Morin transition suppression, Mössbauer spectroscopy was also carried out at 5 K in external magnetic fields up to 10 T as shown in fig. 10, allowing a comparison to observations from *M*(*H*) measurements. The fact that in samples of unoriented particles a random distribution of easy magnetic directions is present, allowing for both spin-flop as well as screw-rotation realignment processes depending on the angle of individual particles' easy axes relative to applied field direction, is also directly evident from spectra shown in fig. 10: At fields up to ca. 2 T – 3 T, only a broadening of the antiferromagnetic spectral component is observed, originating from the superposition of the external field and randomly oriented hyperfine magnetic fields. We reproduced this subspectrum based on a total energy approach derived from that reported by Pankhurst et al.,[56] taking into account contributions to total energy from exchange, anisotropy and magnetic field, which allows for an estimation of the effective anisotropy field $B_A$, based on $B_{sf} \approx \sqrt{2 \cdot B_J \cdot B_A}$ as previously reported for the spin-flop transition field, where $B_J$ is the exchange field.[56] Reproducing the deformation of the AFM spectral component by our fitting routine yields best results for average $B_A \approx 0.009$ T for $p = 1.0$ and $B_A \approx 0.007$ T for particles with $p = 1.25$ (spectra shown for comparison in fig. S8). Using the formula for $B_{sf}$, these values are in general agreement with average spin-flop fields observed in fig. 6 and fig. S10 of ca. 4.8 T and 4.2 T for particle samples with $p = 1.0$ and 1.25. Pankhurst et al. found $B_A \approx 0.02$ T for more bulk-like samples, matching values of $B_{sf}$ in the range of ca. 6.5 T.[56] While determined values of $B_{sf}$ for both samples are decreased in comparison to bulk hematite due to size effects, the further decrease in $B_A$ seems to indicate a continuous reduction in effective magnetic anisotropy in increasingly non-spherical particles, matching the absence of the spin-flop transition in parts of sample $p = 1.5$, shown in fig. S9, and in all particles of higher aspect ratio. To evaluate the progression of the AFM-WFM transition in comparison to $M(H)$ data, WFM fractions extracted from Mössbauer spectra for samples with $p = 1.0$ to $p = 1.5$ are shown in fig. S10. Between 3.5 T and 4.5 T, a deformation in the spectral fine structure appears and points towards the coexistence of the slowly deforming AFM contribution and a second sextet displaying the hyperfine parameters of the WFM state with $2\varepsilon \approx -0.20$ mm/s and $A_{23} \approx 4$. It has to be emphasized that above ca. 6 T, where the spin-flop transition is completed, no further spectral contributions remain except for the narrow WFM sextet. In a previous study van San et al. discuss for powder samples the existence of a critical angle $\varphi_c$ between c-axis and external magnetic field, up to which the spin-flop process takes place, while for higher angles spin realignment is performed via the screw-rotation mechanism.[48] Considering the amount of spectral area with $2\varepsilon$ deviating strongly from $2\varepsilon_{WFM}$ at $B > B_{SF}$ they determined $\varphi_c = 66°$ for ca. 100 nm hematite particles. The absence of such an additional subspectrum for nanospindles discussed here implies a complete WFM alignment around $B_{SF}$ via the spin-flop process independent of the angle $\varphi$, this difference from expected alignment behavior presumably being connected to different anisotropy fields of the nanospindles as compared to particles in Ref. [48].

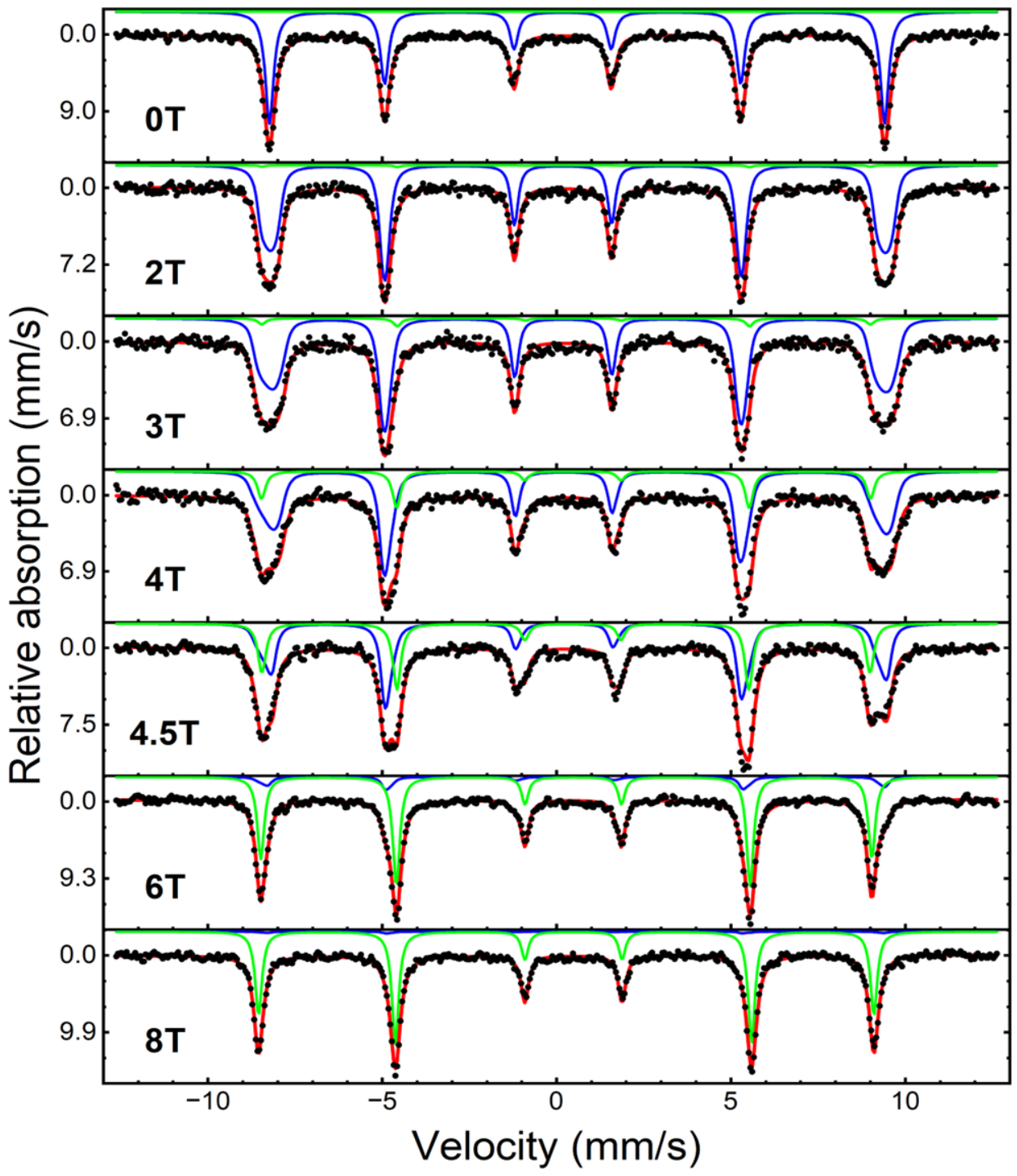

*Figure 10. Mössbauer spectra recorded for p = 1.0 at 5 K and increasing magnetic fields of 0 to 8 T. Spectra comprise of an antiferromagnetic phase (blue), reproduced via a total energy model based on the approach reported by Pankhurst et al.,[56] and a WFM sextet (green). Spectra recorded upon decreasing magnetic field are shown in fig. S7 for comparison.*

To summarize our observations, the existence of a critical aspect ratio for the presence of a temperature- or field-induced AFM-WFM transition for the here-presented hematite nanospindles is clearly and congruently indicated by Mössbauer spectroscopy as well as by (thermo)magnetometry. However, based on the experimental data, the origin of this effect is not immediately apparent. Unlike in ferri- or ferromagnetic elongated nanoparticles, the saturation magnetization $M_S$ of the WFM hematite state is limited to only about 0.33 $Am^2/kg \approx 1800$ A/m (see tab. S1). In a simple approximation the magnetic shape anisotropy of the spindle-shaped particles studied here can be described by that of a prolate ellipsoid as defined by eq. (1).[57]

$$K_{shape} = \frac{1}{2}\mu_0 {M_S}^2 \cdot (N_\perp - N_\parallel) \qquad (1)$$

With $N_\perp \approx 0.472$ and $N_\parallel \approx 0.056$ obtained numerically for an exemplary aspect ratio p of 5, we yield $K_{shape} \approx$ 0.85 $Jm^{-3}$. While comparable, this value is smaller than magnetocrystalline anisotropy (MCA) variations of ca. 1 – 20 $Jm^{-3}$ within the basal plane reported for bulk hematite,[50] whereby it seems insufficient as an immediate cause to facilitate the effects observed here, especially as it would lead to minimum energy for magnetization along the long particle axis, inconsistent with observations for the nanospindles. Furthermore, one has to consider that the observed suppression of the Morin transition already takes place at far lower aspect ratios close to p = 1.5, corresponding to even lower values of shape anisotropy.

At the same time, neither XRD nor TEM show any considerable trend in crystallite size or dimensions of the primary particle subunits in our samples, which could influence the Morin transition. In previous studies on hematite nanostructures of varying shape, authors discussed different mechanisms like defect concentration, deviating lattice constants and their variation by strain as well as surface and shape anisotropy effects as possible causes for decreasing $T_{Morin}$.[21, 58] Also, Hill et al. reported the absence of the Morin transition in mesoporous samples, also affecting hematite lattice parameters.[59] Wang et al. proposed a suppression based on the existence of surface magnetic anisotropies of different sign for different surface facets in nanorings and nanocylinders, respectively, resulting in a shift of the sensitive anisotropy energy equilibrium depending on the area of the respective (001) and (111) surface facets.[35] This would be compatible with trends of the nanospindles, increasing the relative fraction of facets perpendicular to (111) proportionally to the aspect ratio. The importance of crystallographic orientation of surface facets for the shape induced anisotropy of antiferromagnetic nanoparticles was discussed by Gomonay et al. based on theoretical calculations,[60] whereas Chen et al. and Xu et al. experimentally found strongly varying remanence and coercivity by comparing several hematite nanostructures, comparable in size to the spindle particles discussed here, dependent on the exposed surface facets.[61, 62] While we cannot provide concrete evidence that this is the driving mechanism for Morin transition suppression in our nanostructured hematite spindle particles, these findings provide validity to the hypothesis that higher aspect ratios of the hematite nanospindles by elongation along the c-axis can shift the sensitive equilibrium of magnetic anisotropy contributions towards a stabilization of the WFM spin structure. Thereby, this mechanism would provide a valuable addition to the toolbox available to tune the behavior of future nanoparticles: It establishes the spindle elongation as a further parameter to tune $T_{Morin}$ independent of the particle size and enables the conservation of the WFM hematite state for much higher particle dimensions than usually assumed possible, e.g. for magnetically controllable particles in complex materials.

**Conclusion**

The correlation between aspect ratio and magnetic behavior is investigated for hematite nanospindles with a short axis length of about 70 nm and aspect ratios *p* between 1 and 5.2, prepared following a hydrothermal approach. The particle formation process is governed by epitaxial merging of primarily formed particles of about 30 nm diameter, as estimated from TEM data, accompanied by the presence of nm-sized nanopores. Our findings reveal that, for samples with an aspect ratio exceeding 1.5, the Morin transition is fully suppressed down to cryogenic temperatures. This result is surprising, given that the aspect ratio is proportional to the particle volume, and previous studies in the literature suggest that the suppression of the antiferromagnetic state is less pronounced upon increasing particle volume.

Studying the degree of suppression in more detail in ZFC-FC magnetometry experiments, starting from $T_{Morin} \approx 200$ K for $p = 1.0$, a continuous decrease in $T_{Morin}$ and broadening of thermal hysteresis was found with increasing $p$, with $p = 1.5$ being identified as close to a critical aspect ratio. This observation is corroborated by Mössbauer spectroscopy data, indicating that for $p = 1.5$ about 20 % of the material no longer experience the Morin transition. This approach also reveals the presence of spin misalignment in the WFM state for this critical aspect ratio, as evident from deviations in the nuclear quadrupole level shift.

Comparing $M(H)$- and in-field Mössbauer spectroscopy data allows us to discern contributions of alignment and WFM/AFM phase composition and net magnetization. Data analysis indicates a decrease in effective magnetic anisotropy towards the critical aspect ratio of $p \approx 1.5$, in agreement with the trend in $T_{Morin}$ observed in temperature dependent experiments. The combined data obtained in our study thus demonstrate the aspect ratio to be of critical importance when discussing the magnetic behavior of hematite nanoparticles, and their unexpected ability to conserve the WFM state in much larger particles than generally anticipated.

**Supporting Information:**

Supplementary Information can be found under (insert DOI), including: Further TEM analysis, information on the dependence of nanospindle elongation on synthesis conditions, temperature-dependence of the field-induced AFM-WFM phase transition measured via magnetometry, nanospindle magnetic parameters from magnetometry at 4.3 and 300 K, detailed Mössbauer spectroscopy study and corresponding hyperfine parameters across the Morin transition.

**Acknowledgements:**

Support by the Interdisciplinary Center for Analytics on the Nanoscale (ICAN) of the University of Duisburg-Essen (DFG RIsources reference: RI_00313), a DFG-registered core facility (Project Nos. 233512597 and 324659309), is gratefully acknowledged

This work was supported by the DFG via projects LA5175/1-1 and SCHM1747/16-1 and CRC/TRR 270 (subproject B05) project ID 405553726.

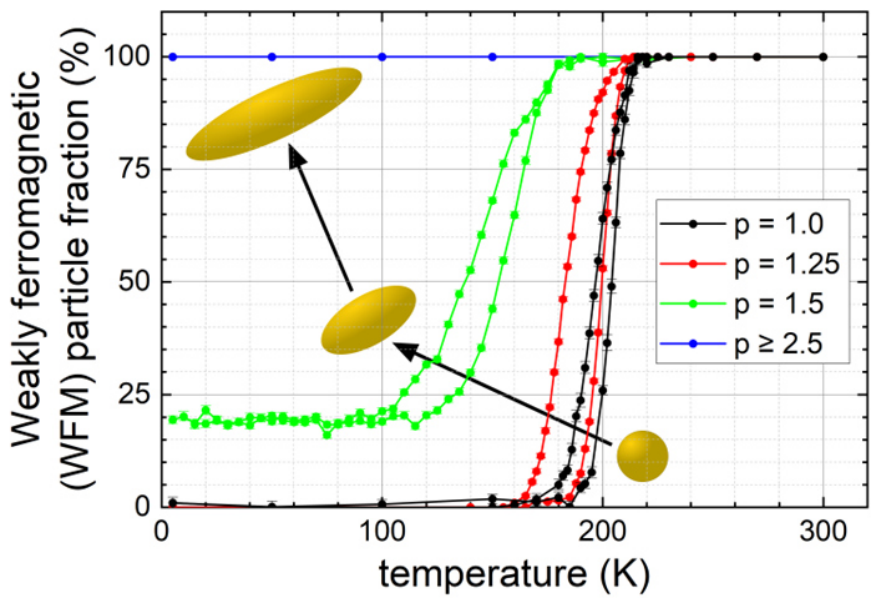

Table of Contents Image

References:

(1) Chourpa, I.; Douziech-Eyrolles, L.; Ngaboni-Okassa, L.; Fouquenet, J. F.; Cohen-Jonathan, S.; Souce, M.; Marchais, H.; Dubois, P. Molecular Composition of Iron Oxide Nanoparticles, Precursors for Magnetic Drug Targeting, As Characterized by Confocal Raman Microspectroscopy. *Analyst* **2005**, *130*, 1395-1403. DOI: 10.1039/b419004a
(2) Jain, T. K.; Morales, M. A.; Sahoo, S. K.; Leslie-Pelecky, D. L.; Labhasetwar, V. Iron Oxide Nanoparticles for Sustained Delivery of Anticancer Agents. *Mol Pharm* **2005**, *2*, 194-205. DOI: 10.1021/mp0500014
(3) Xie, J.; Liu, G.; Eden, H. S.; Ai, H.; Chen, X. Surface-Engineered Magnetic Nanoparticle Platforms for Cancer Imaging and Therapy. *Acc Chem Res* **2011**, *44*, 883-892. DOI: 10.1021/ar200044b
(4) Yu, M. K.; Park, J.; Jon, S. Targeting Strategies for Multifunctional Nanoparticles in Cancer Imaging and Therapy. *Theranostics* **2012**, *2*, 3-44. DOI: 10.7150/thno.3463
(5) Modo, M.; Hoehn, M.; Bulte, J. W. Cellular MR Imaging. *Mol Imaging* **2005**, *4*, 143-164. DOI: 10.1162/15353500200505145

(6) Boutry, S.; Laurent, S.; Elst, L. V.; Muller, R. N. Specific E-selectin Targeting With a Superparamagnetic MRI Contrast Agent. *Contrast Media Mol Imaging* **2006**, *1*, 15-22. DOI: 10.1002/cmmi.87
(7) Bulte, J. W. Intracellular Endosomal Magnetic Labeling of Cells. *Methods Mol Med* **2006**, *124*, 419-439. DOI: 10.1385/1-59745-010-3:419
(8) Sun, S.; Murray, C. B.; Weller, D.; Folks, L.; Moser, A. Monodisperse FePt Nanoparticles and Ferromagnetic FePt Nanocrystal Superlattices. *Science* **2000**, *287*, 1989-1992. DOI: 10.1126/science.287.5460.1989
(9) Wagner, J.; Autenrieth, T.; Hempelmann, R. Core Shell Particles Consisting of Cobalt Ferrite and Silica As Model Ferrofluids [$CoFe_2O_4$-$SiO_2$ Core Shell Particles]. *Journal of Magnetism and Magnetic Materials* **2002**, *252*, 4-6. DOI: 10.1016/S0304-8853(02)00729-1
(10) Ren, B.; Ruditskiy, A.; Song, J. H.; Kretzschmar, I. Assembly Behavior of Iron Oxide-Capped Janus Particles in a Magnetic Field. *Langmuir* **2012**, *28*, 1149-1156. DOI: 10.1021/la203969f
(11) Sinn, I.; Kinnunen, P.; Pei, S. N.; Clarke, R.; McNaughton, B. H.; Kopelman, R. Magnetically Uniform and Tunable Janus Particles. *Appl Phys Lett* **2011**, *98*, 024101-024101. DOI: 10.1063/1.3541876
(12) Dutz, S.; Kettering, M.; Hilger, I.; Muller, R.; Zeisberger, M. Magnetic Multicore Nanoparticles for Hyperthermia–Influence of Particle Immobilization in Tumour Tissue on Magnetic Properties. *Nanotechnology* **2011**, *22*, 265102. DOI: 10.1088/0957-4484/22/26/265102
(13) Garcia-Soriano, D.; Milan-Rois, P.; Lafuente-Gomez, N.; Rodriguez-Diaz, C.; Navio, C.; Somoza, A.; Salas, G. Multicore Iron Oxide Nanoparticles for Magnetic Hyperthermia and Combination Therapy Against Cancer Cells. *J Colloid Interface Sci* **2024**, *670*, 73-85. DOI: 10.1016/j.jcis.2024.05.046
(14) Yoon, T. J.; Lee, H.; Shao, H. L.; Hilderbrand, S. A.; Weissleder, R. Multicore Assemblies Potentiate Magnetic Properties of Biomagnetic Nanoparticles. *Adv Mater* **2011**, *23*, 4793-4797. DOI: 10.1002/adma.201102948
(15) Koch, K.; Kundt, M.; Eremin, A.; Nadasi, H.; Schmidt, A. M. Efficient Ferronematic Coupling With Polymer-Brush Particles. *Phys Chem Chem Phys* **2020**, *22*, 2087-2097. DOI: 10.1039/c9cp06245a
(16) Seifert, J.; Roitsch, S.; Schmidt, A. M. Covalent Hybrid Elastomers Based on Anisotropic Magnetic Nanoparticles and Elastic Polymers. *Acs Appl Polym Mater* **2021**, *3*, 1324-1337. DOI: 10.1021/acsapm.0c00950
(17) Seifert, J.; Koch, K.; Hess, M.; Schmidt, A. M. Magneto-Mechanical Coupling of Single Domain Particles in Soft Matter Systems. *Phys Sci Rev* **2022**, *7*, 1237-1261. DOI: 10.1515/psr-2019-0092
(18) Eberbeck, D.; Gustafsson, S.; Olsson, E.; Braun, K. F.; Gollwitzer, C.; Krumrey, M.; Bergemann, C.; Wang, A.; Yu, W. W.; Kratz, H.; Hankiewicz, B.; Messing, R.; Steffens, N.; Schmidt, A. M.; Schmidt, C.; Müller, R.; Wiekhorst, F. Magneto-Structural Characterization of Different Kinds of Magnetic Nanoparticles. *Journal of Magnetism and Magnetic Materials* **2023**, *583*, 171031. DOI: 10.1016/j.jmmm.2023.171031
(19) Do, H. M.; Le, T. H. P.; Tran, D. T.; Nguyen, T. N. A.; Skorvanek, I.; Kovac, J.; Svec, P. J. r.; Phan, M. H. Magnetic Interaction Effects in $Fe_3O_4$@$CoFe_2O_4$ Core/Shell Nanoparticles. *J Sci-Adv Mater Dev* **2024**, *9*, 100658. DOI: 10.1016/j.jsamd.2023.100658
(20) Jia, C. J.; Sun, L. D.; Luo, F.; Han, X. D.; Heyderman, L. J.; Yan, Z. G.; Yan, C. H.; Zheng, K.; Zhang, Z.; Takano, M.; Hayashi, N.; Eltschka, M.; Kläui, M.; Rüdiger, U.; Kasama, T.; Cervera-Gontard, L.; Dunin-Borkowski, R. E.; Tzvetkov, G.; Raabe, J. Large-Scale Synthesis of Single-Crystalline Iron Oxide Magnetic Nanorings. *J Am Chem Soc* **2008**, *130*, 16968-16977. DOI: 10.1021/ja805152t
(21) Mitra, S.; Das, S.; Basu, S.; Sahu, P.; Mandal, K. Shape- and Field-Dependent Morin Transitions in Structured α-$Fe_2O_3$. *Journal of Magnetism and Magnetic Materials* **2009**, *321*, 2925-2931. DOI: 10.1016/j.jmmm.2009.04.044
(22) Supattarasakda, K.; Petcharoen, K.; Permpool, T.; Sirivat, A.; Lerdwijitjarud, W. Control of Hematite Nanoparticle Size and Shape by the Chemical Precipitation Method. *Powder Technol* **2013**, *249*, 353-359. DOI: 10.1016/j.powtec.2013.08.042
(23) Dzyaloshinsky, I. A Thermodynamic Theory of Weak Ferromagnetism of Antiferromagnetics. *J Phys Chem Solids* **1958**, *4*, 241-255. DOI: 10.1016/0022-3697(58)90076-3
(24) Flanders, P. J.; Remeika, J. P. Magnetic Properties of Hematite Single Crystals. *The Philosophical Magazine: A Journal of Theoretical Experimental and Applied Physics* **1965**, *11*, 1271-1288. DOI: 10.1080/14786436508224935
(25) Cornell, R. M.; Schwertmann, U. *The Iron Oxides: Structure, Properties, Reactions, Occurrences, and Uses*; Wiley-VCH, 2003.
(26) Moriya, T. New Mechanism of Anisotropic Superexchange Interaction. *Phys Rev Lett* **1960**, *4*, 228-230. DOI: 10.1103/PhysRevLett.4.228
(27) Ruskov, T.; Tomov, T.; Georgiev, S. Mössbauer Investigation of Morin Transition in Hematite. *Phys Status Solidi A* **1976**, *37*, 295-302. DOI: 10.1002/pssa.2210370137
(28) Morin, F. J. Magnetic Susceptibility of α-$Fe_2O_3$ and α-$Fe_2O_3$ With Added Titanium. *Physical Review* **1950**, *78*, 819-820. DOI: 10.1103/PhysRev.78.819.2
(29) Anderson, P. W.; Merritt, F. R.; Remeika, J. P.; Yager, W. A. Magnetic Resonance in α-$Fe_2O_3$. *Physical Review* **1954**, *93*, 717-718. DOI: 10.1103/PhysRev.93.717
(30) Shimomura, N.; Pati, S. P.; Sato, Y.; Nozaki, T.; Shibata, T.; Mibu, K.; Sahashi, M. Morin Transition Temperature in (0001)-Oriented α-$Fe_2O_3$ Thin Film and Effect of Ir Doping. *J Appl Phys* **2015**, *117*, 17C736. DOI: 10.1063/1.4916304
(31) Özdemir, Ö.; Dunlop, D. J.; Berquó, T. S. Morin Transition in Hematite: Size Dependence and Thermal Hysteresis. *Geochem Geophy Geosy* **2008**, *9*. DOI: 10.1029/2008gc002110
(32) Bengoa, J. F.; Alvarez, A. M.; Bianchi, A. E.; Punte, G.; Vandenberghe, R. E.; Mercader, R. C.; Marchetti, S. G. The Morin Transition in Nanostructured Pseudocubic Hematite: Effect of the Intercrystallite Magnetic Exchange. *Mater Chem Phys* **2010**, *123*, 191-198. DOI: 10.1016/j.matchemphys.2010.03.081
(33) Kubániová, D.; Kubícková, L.; Kmjec, T.; Záveta, K.; Niznansky, D.; Brázda, P.; Klementová, M.; Kohout, J. Hematite: Morin Temperature of Nanoparticles With Different Size. *Journal of Magnetism and Magnetic Materials* **2019**, *475*, 611-619. DOI: 10.1016/j.jmmm.2018.11.126
(34) Chuev, M. A.; Mishchenko, I. N.; Kubrin, S. P.; Lastovina, T. A. Novel Insight Into the Effect of Disappearance of the Morin Transition in Hematite Nanoparticles. *Jetp Lett+* **2017**, *105*, 700-705. DOI: 10.1134/S0021364017110042
(35) Wang, J.; Aguilar, V.; Li, L.; Li, F. G.; Wang, W. Z.; Zhao, G. M. Strong Shape-Dependence of Morin Transition in α-$Fe_2O_3$ Single-Crystalline Nanostructures. *Nano Res* **2015**, *8*, 1906-1916. DOI: 10.1007/s12274-014-0700-z
(36) Siddiqui, S. A.; Hong, D. S.; Pearson, J. E.; Hoffmann, A. Antiferromagnetic Oxide Thin Films for Spintronic Applications. *Coatings* **2021**, *11*. DOI: 10.3390/coatings11070786
(37) Pauling, L.; Hendricks, S. B. The Crystal Structures of Hematite and Corundum. *J Am Chem Soc* **1925**, *47*, 781-790. DOI: 10.1021/ja01680a027

(38) Artman, J. O.; Murphy, J. C.; Foner, S. Magnetic Anisotropy in Antiferromagnetic Corundum-Type Sesquioxides. *Physical Review* **1965**, *138*, A912-A917. DOI: 10.1103/PhysRev.138.A912
(39) Ocana, M.; Morales, M. P.; Serna, C. J. The Growth-Mechanism of α-$Fe_2O_3$ Ellipsoidal Particles in Solution. *J Colloid Interf Sci* **1995**, *171*, 85-91. DOI: 10.1006/jcis.1995.1153
(40) Ozaki, M.; Kratohvil, S.; Matijevic, E. Formation of Monodispersed Spindle-Type Hematite Particles. *J Colloid Interf Sci* **1984**, *102*, 146-151. DOI: 10.1016/0021-9797(84)90208-X
(41) Jones, D. H.; Srivastava, K. K. P. Many-State Relaxation Model for the Mössbauer-Spectra of Superparamagnets. *Phys Rev B* **1986**, *34*, 7542-7548. DOI: 10.1103/PhysRevB.34.7542
(42) Szymański, K.; Dobrzyński, L. Simple Approximation to the Transmission Integral in Mössbauer Spectroscopy. *Nuclear Instruments and Methods in Physics Research Section B: Beam Interactions with Materials and Atoms* **1990**, *51*, 192-195. DOI: 10.1016/0168-583X(90)90522-V
(43) Zhu, M. Y.; Wang, Y.; Meng, D. H.; Qin, X. Z.; Diao, G. W. Hydrothermal Synthesis of Hematite Nanoparticles and Their Electrochemical Properties. *J Phys Chem C* **2012**, *116*, 16276-16285. DOI: 10.1021/jp304041m
(44) Blake, R. L.; Hessevick, R. E.; Zoltai, T.; Finger, L. W. Refinement of Hematite Structure. *Am Mineral* **1966**, *51*, 123-129.
(45) Porath, H. Stress Induced Magnetic Anisotropy in Natural Single Crystals of Hematite. *Philos Mag* **1968**, *17*, 603-608. DOI: 10.1080/14786436808217746
(46) Shapira, Y. Ultrasonic Behavior Near Spin-Flop Transitions of Hematite. *Physical Review* **1969**, *184*, 589-600. DOI: 10.1103/PhysRev.184.589
(47) Bakkaloglu, O. F.; Pankhurst, Q. A.; Thomas, M. F. Field-Induced Phase Transitions in Neutron-Irradiated Haematite. *Journal of Physics: Condensed Matter* **1993**, *5*, 3265. DOI: 10.1088/0953-8984/5/19/022
(48) Van San, E.; De Grave, E.; Vandenberghe, R. E. Field-Induced Spin Transitions in Hematite Powders As Observed From Mössbauer Spectroscopy. *Journal of Magnetism and Magnetic Materials* **2004**, *269*, 54-60. DOI: 10.1016/S0304-8853(03)00561-4
(49) Zysler, R. D.; Fiorani, D.; Testa, A. M.; Suber, L.; Agostinelli, E.; Godinho, M. Size Dependence of the Spin-Flop Transition in Hematite Nanoparticles. *Phys Rev B* **2003**, *68*, 212408. DOI: 10.1103/PhysRevB.68.212408
(50) Flanders, P. J.; Schuele, W. J. Anisotropy in the Basal Plane of Hematite Single Crystals. *The Philosophical Magazine: A Journal of Theoretical Experimental and Applied Physics* **1964**, *9*, 485-490. DOI: 10.1080/14786436408222959
(51) Martin-Hernandez, F.; Guerrero-Suárez, S. Magnetic Anisotropy of Hematite Natural Crystals: High Field Experiments. *Int J Earth Sci* **2012**, *101*, 637-647. DOI: 10.1007/s00531-011-0665-z
(52) Kündig, W.; Bommel, H.; Constaba.G; Lindqui.Rh. Some Properties of Supported Small α-$Fe_2O_3$ Particles Determined With Mössbauer Effect. *Physical Review* **1966**, *142*, 327-333. DOI: 10.1103/PhysRev.142.327
(53) Reufer, M.; Dietsch, H.; Gasser, U.; Grobety, B.; Hirt, A. M.; Malik, V. K.; Schurtenberger, P. Magnetic Properties of Silica Coated Spindle-Type Hematite Particles. *J Phys-Condens Mat* **2011**, *23*, 065102. DOI: 10.1088/0953-8984/23/6/065102
(54) Suber, L.; Imperatori, P.; Mari, A.; Marchegiani, G.; Mansilla, M. V.; Fiorani, D.; Plunkett, W. R.; Rinaldi, D.; Cannas, C.; Ennas, G.; Peddis, D. Thermal Hysteresis of Morin Transition in Hematite Particles. *Phys Chem Chem Phys* **2010**, *12*, 6984-6989. DOI: 10.1039/b925371h
(55) Morup, S.; Topsoe, H. Mössbauer Studies of Thermal Excitations in Magnetically Ordered Microcrystals. *Appl Phys* **1976**, *11*, 63-66. DOI: Doi 10.1007/Bf00895017

(56) Pankhurst, Q. A.; Pollard, R. J. Mössbauer-Spectra of Antiferromagnetic Powders in Applied Fields. *J Phys-Condens Mat* **1990**, *2*, 7329-7337. DOI: 10.1088/0953-8984/2/35/008
(57) Failde, D.; Ocampo-Zalvide, V.; Serantes, D.; Iglesias, O. Understanding Magnetic Hyperthermia Performance Within the "Brezovich Criterion": Beyond the Uniaxial Anisotropy Description. *Nanoscale* **2024**, *16*, 14319-14329. DOI: 10.1039/d4nr02045f
(58) Liu, Q. S.; Barrón, V.; Torrent, J.; Qin, H. F.; Yu, Y. The Magnetism of Micro-sized Hematite Explained. *Phys Earth Planet In* **2010**, *183*, 387-397. DOI: 10.1016/j.pepi.2010.08.008
(59) Hill, A. H.; Jiao, F.; Bruce, P. G.; Harrison, A.; Kockelmann, W.; Ritter, C. Neutron Diffraction Study of Mesoporous and Bulk Hematite, α-$Fe_2O_3$. *Chem Mater* **2008**, *20*, 4891-4899. DOI: 10.1021/cm800009s
(60) Gomonay, O.; Kondovych, S.; Loktev, V. Shape-Induced Anisotropy in Antiferromagnetic Nanoparticles. *Journal of Magnetism and Magnetic Materials* **2014**, *354*, 125-135. DOI: 10.1016/j.jmmm.2013.11.003
(61) Chen, J. S.; Chen, X.; Li, Y. M.; Chen, X. L.; Ramanujan, R. V.; Hu, X. Distinct optical and magnetic properties of ionic liquid tuned hematite nanocrystals having different exposed (001) facets. *Rsc Adv* **2014**, *4*, 593-597. DOI: 10.1039/c3ra43924k
(62) Xu, G.; Li, L. L.; Shen, Z. J.; Tao, Z. H.; Zhang, Y.; Tian, H.; Wei, X.; Shen, G.; Han, G. R. Magnetite $Fe_3O_4$ nanoparticles and hematite α-$Fe_2O_3$ uniform oblique hexagonal microdisks, drum-like particles and spindles and their magnetic properties. *J Alloy Compd* **2015**, *629*, 36-42. DOI: 10.1016/j.jallcom.2014.11.140

Supplementary Information

# Insight into the Correlation of Shape and Magnetism of Hematite Nanospindles

*Juri Kopp[1], Gerald Richwien[2], Markus Heidelmann[3], Benoît Rhein[2], Soma Salamon[1], Annette M. Schmidt[2], Joachim Landers[1*]*

[1] Faculty of Physics and Center for Nanointegration Duisburg-Essen (CENIDE), University of Duisburg-Essen, 47057, Duisburg, Germany

[2] Department of Chemistry & Biochemistry, University of Cologne, 50939 Cologne, Germany

[3] ICAN, Interdisciplinary Center for Analytics on the Nanoscale, University of Duisburg-Essen, 47057, Duisburg, Germany

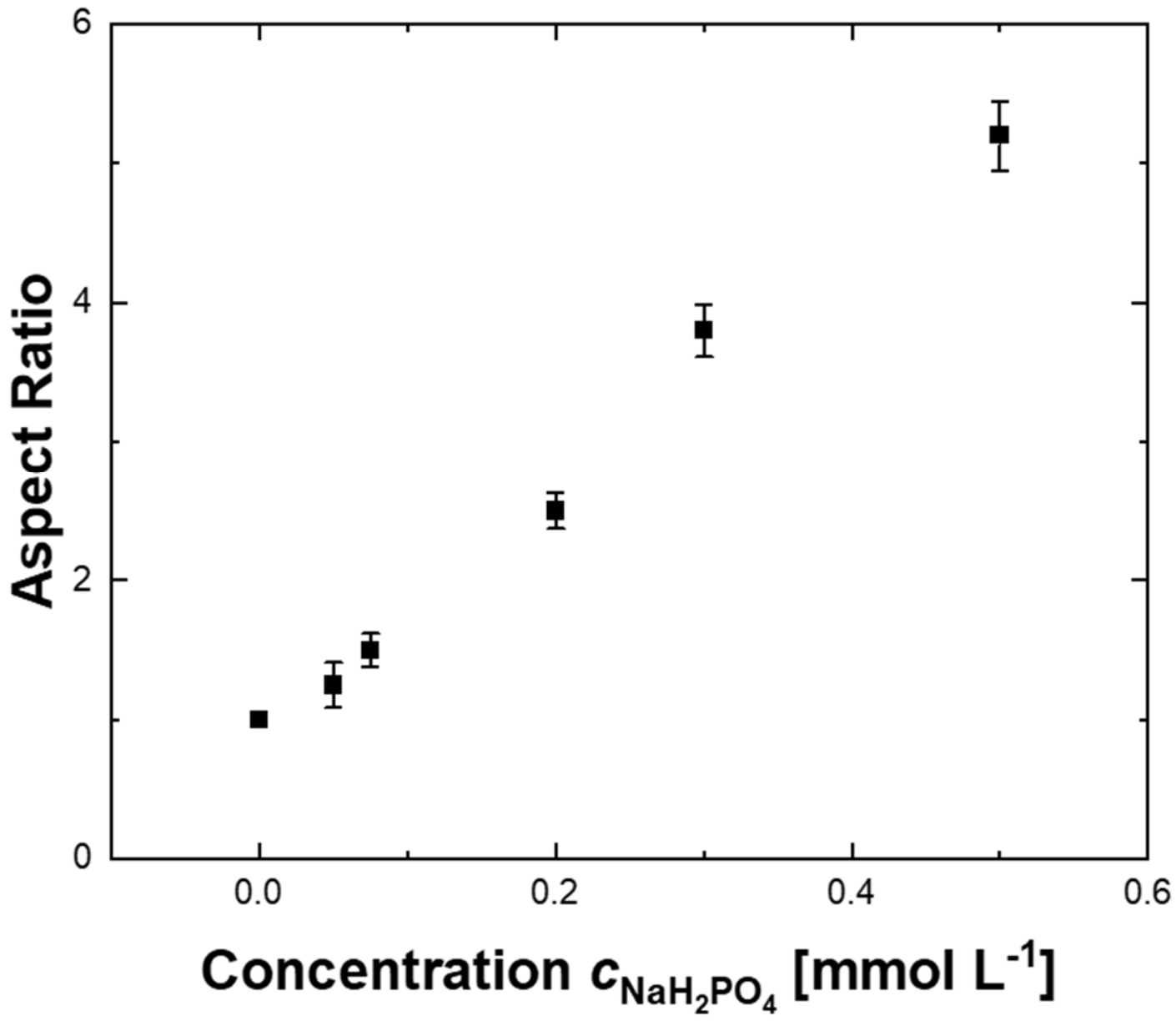

*Figure S1: Aspect ratio of spindle-shaped nanoparticles extracted from TEM data vs $NaH_2PO_4$ concentration added during synthesis.*

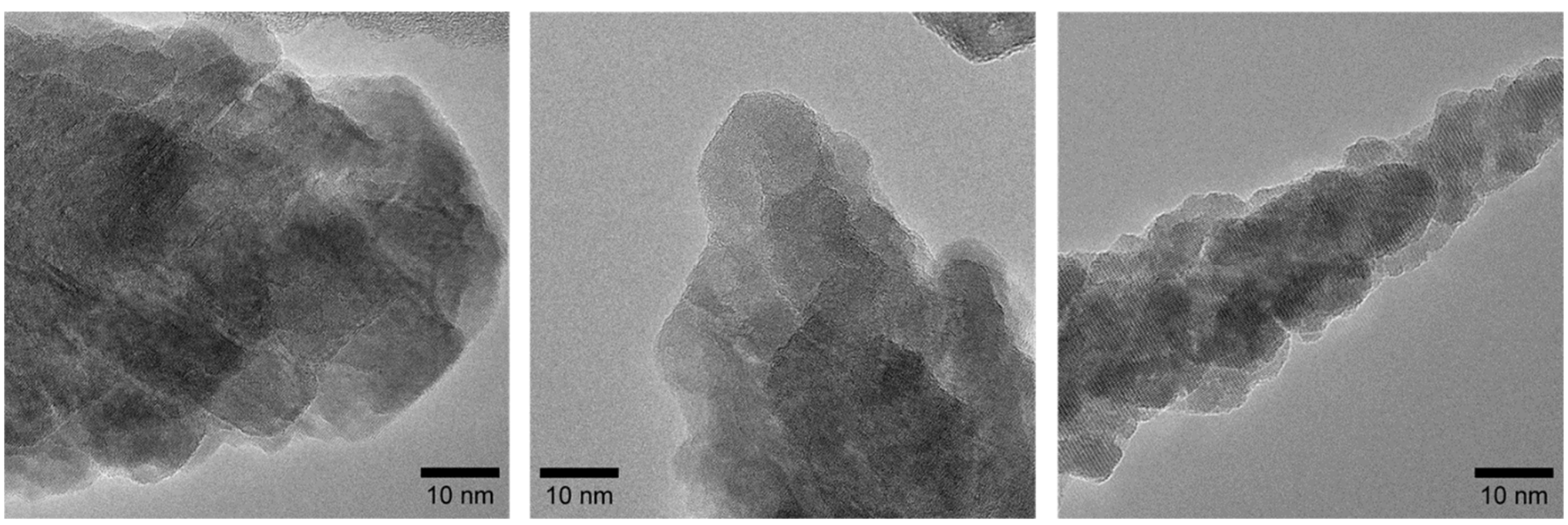

*Figure S2: TEM images of spindle-shaped nanoparticles from left to right with aspect ratios p of 2.5, 3.8, and 5.2. Shown are the edges of individual nanospindles to illustrate shape and size of partially merged former primary particles.*

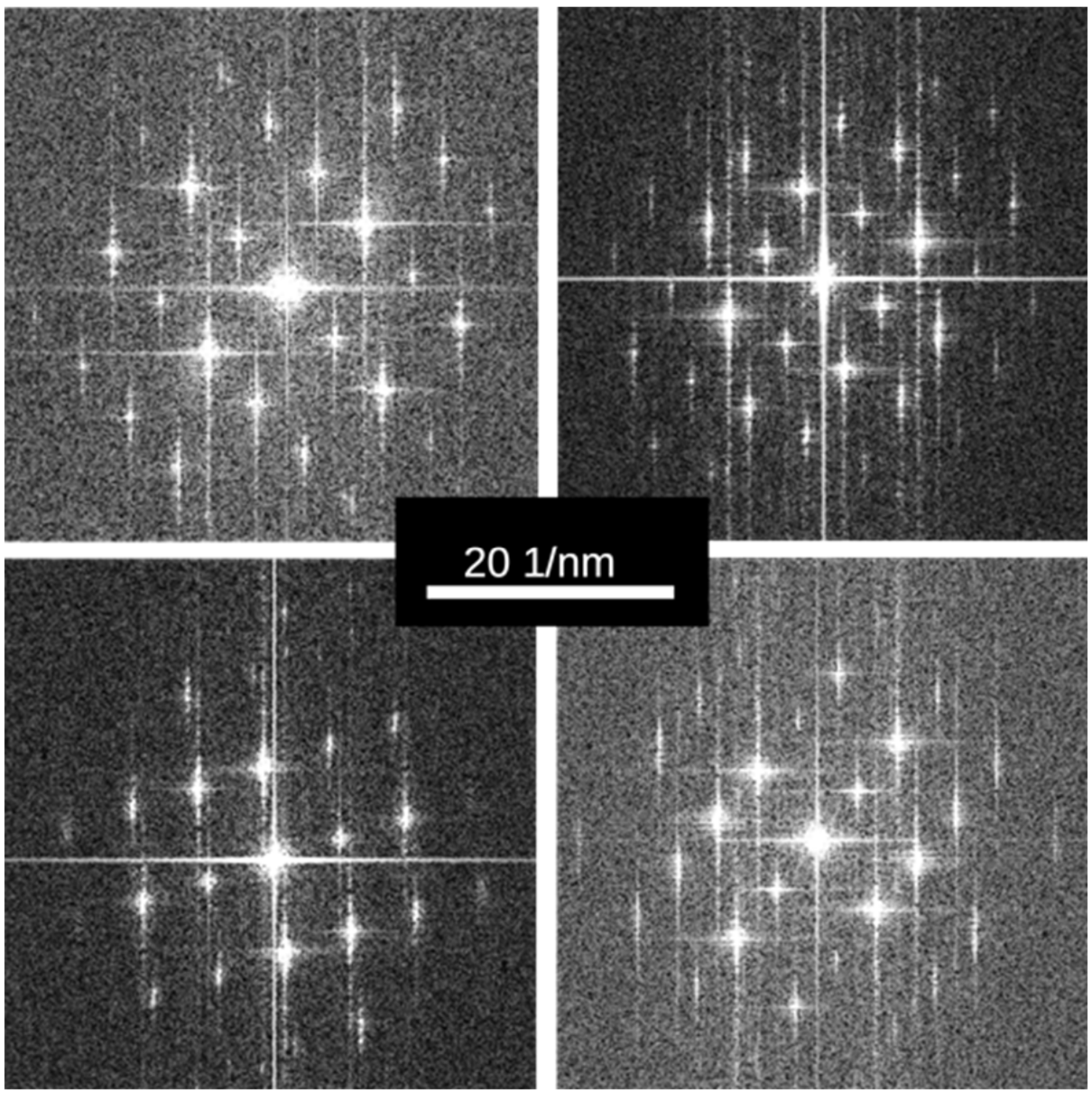

*Figure S3: Fourier transformed versions of the images in fig. 4.*

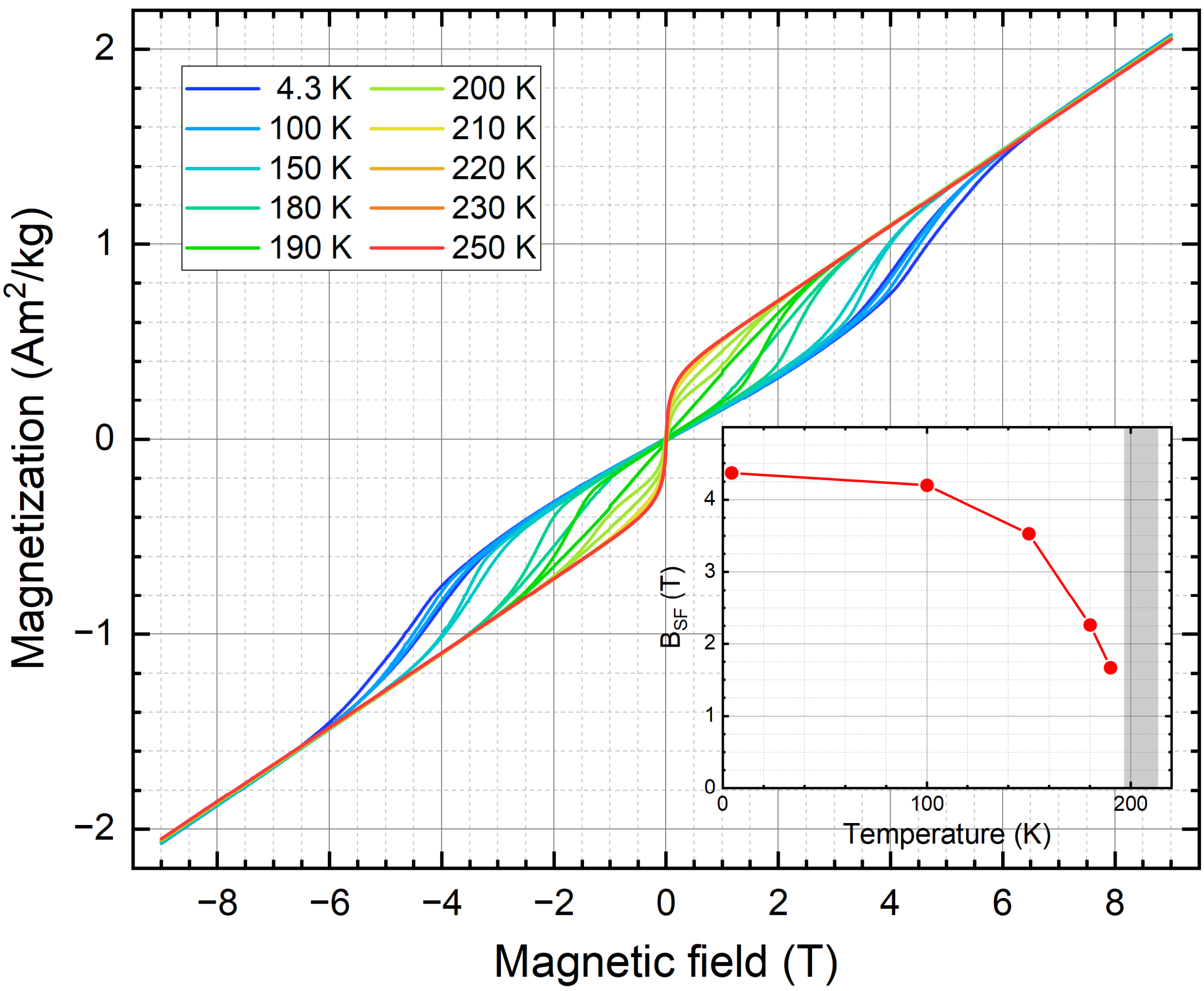

*Figure S4: M(H) curves for aspect ratio p = 1.0 recorded between 4.3 K and 250 K showing the field-induced AFM-WFM transition. The inset shows spin-flop fields $B_{SF}$, defined here by the inflection points of magnetization upon rising magnetic field, versus temperature, the gray area highlights the temperature region of thermal hysteresis for this sample.*

*Tab. S1: Coercivity $H_C$, WFM saturation magnetization $M_S$, remanent magnetization $M_R$ and magnetic mass susceptibility $\chi_{WFM}$ in the WFM state measured at 4.3 K and 300 K for nanospindles of aspect ratio p = 1.0 to 5.2. $M_S$ and $\chi_{WFM}$ were obtained from interpolation of the high-field region.*

| T (K) | p | $H_C$ (mT) | $M_S$ (Am²/kg) | $M_R$ (Am²/kg) | $\chi_{WFM}$ (Am²/kgOe) |
|---|---|---|---|---|---|
| 4.3 | 1.0 | 39 | 0.325 | 0.007 | $1.99 \cdot 10^{-5}$ |
| 4.3 | 1.25 | 28 | 0.330 | 0.007 | $2.02 \cdot 10^{-5}$ |
| 4.3 | 1.5 | 248 | 0.322 | 0.064 | $1.90 \cdot 10^{-5}$ |
| 4.3 | 2.5 | 174 | 0.342 | 0.205 | $1.91 \cdot 10^{-5}$ |
| 4.3 | 3.8 | 195 | 0.327 | 0.203 | $2.12 \cdot 10^{-5}$ |
| 4.3 | 5.2 | 214 | 0.319 | 0.184 | $2.14 \cdot 10^{-5}$ |
| 300 | 1.0 | 0 | 0.326 | 0 | $1.95 \cdot 10^{-5}$ |
| 300 | 1.25 | 4 | 0.332 | 0.034 | $1.95 \cdot 10^{-5}$ |
| 300 | 1.5 | 2 | 0.316 | 0.014 | $1.83 \cdot 10^{-5}$ |
| 300 | 2.5 | 6 | 0.338 | 0.041 | $1.84 \cdot 10^{-5}$ |
| 300 | 3.8 | 18 | 0.326 | 0.079 | $2.05 \cdot 10^{-5}$ |
| 300 | 5.2 | 21 | 0.315 | 0.067 | $2.03 \cdot 10^{-5}$ |

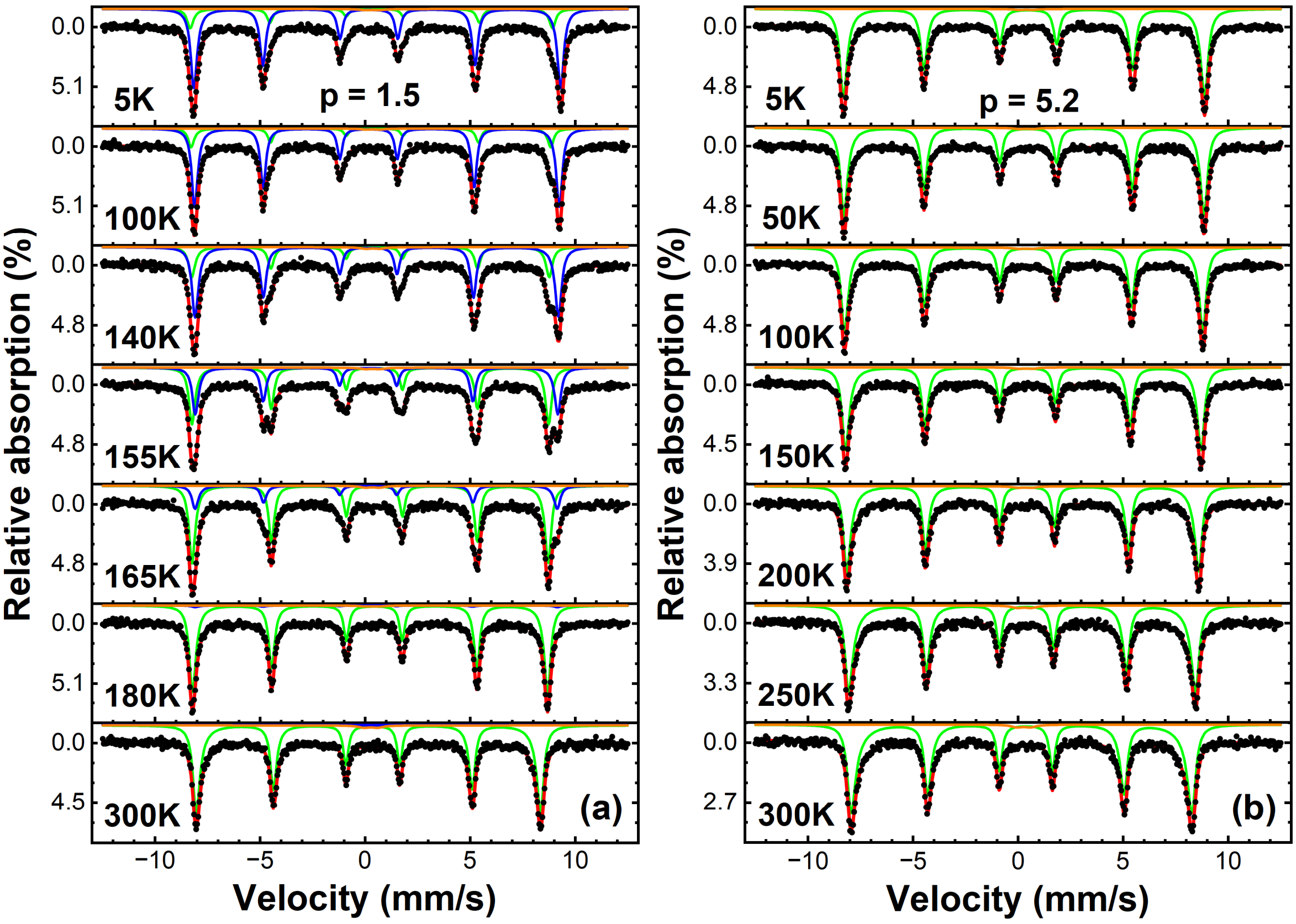

*Figure S5: Mössbauer spectra recorded from 5 K to 300 K, showing the suppression of the Morin transition in particles of aspect ratio 5.2 (b) and the partially suppressed transition shifted to lower temperatures for aspect ratio 1.5 (a).*

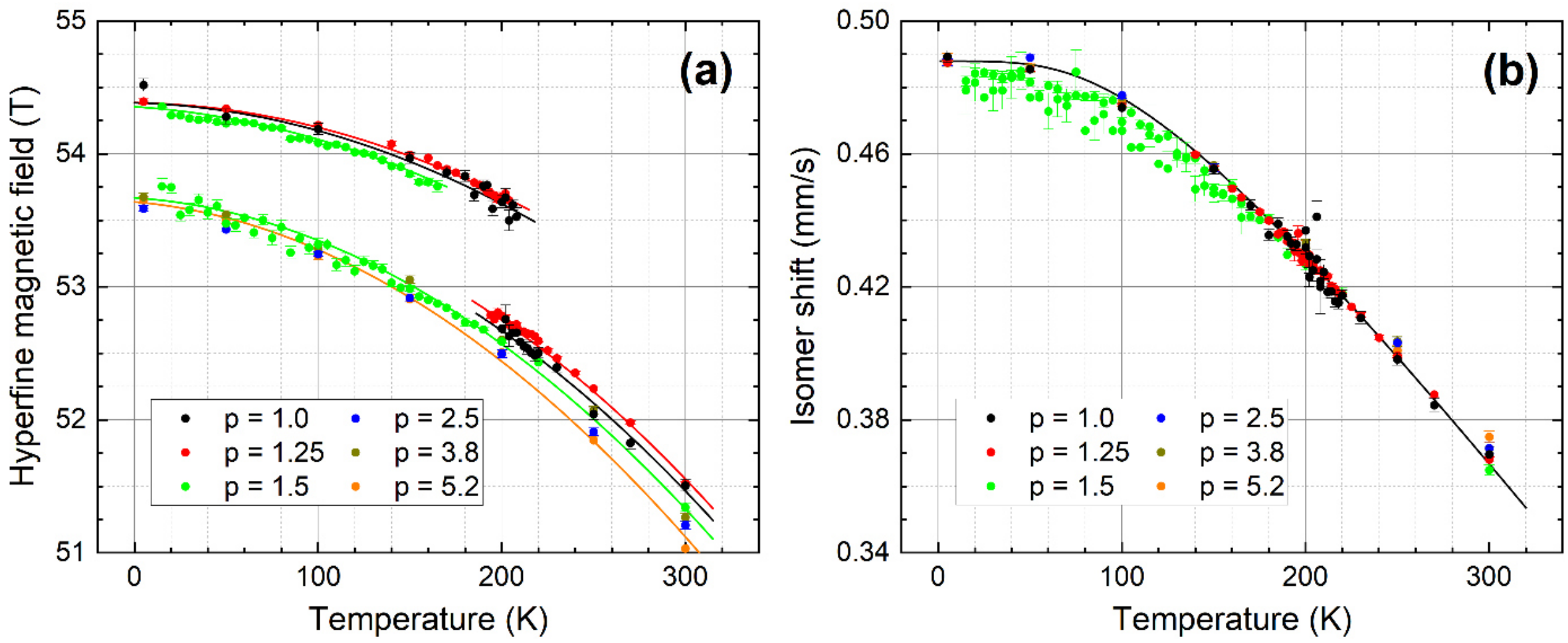

*Figure S6: Hyperfine magnetic field $B_{HF}$ (a) and isomer shift $\delta$ (b) extracted from Mössbauer spectra recorded upon heating from 5 K to 300 K for p = 1.0 - 5.2. Minor differences for $B_{HF}$ approaching 300 K are assigned to slightly different degrees of superparamagnetic relaxation. Solid lines are guide-to-the-eye, isomer shifts are given relative to α-Fe at ambient temperature.*

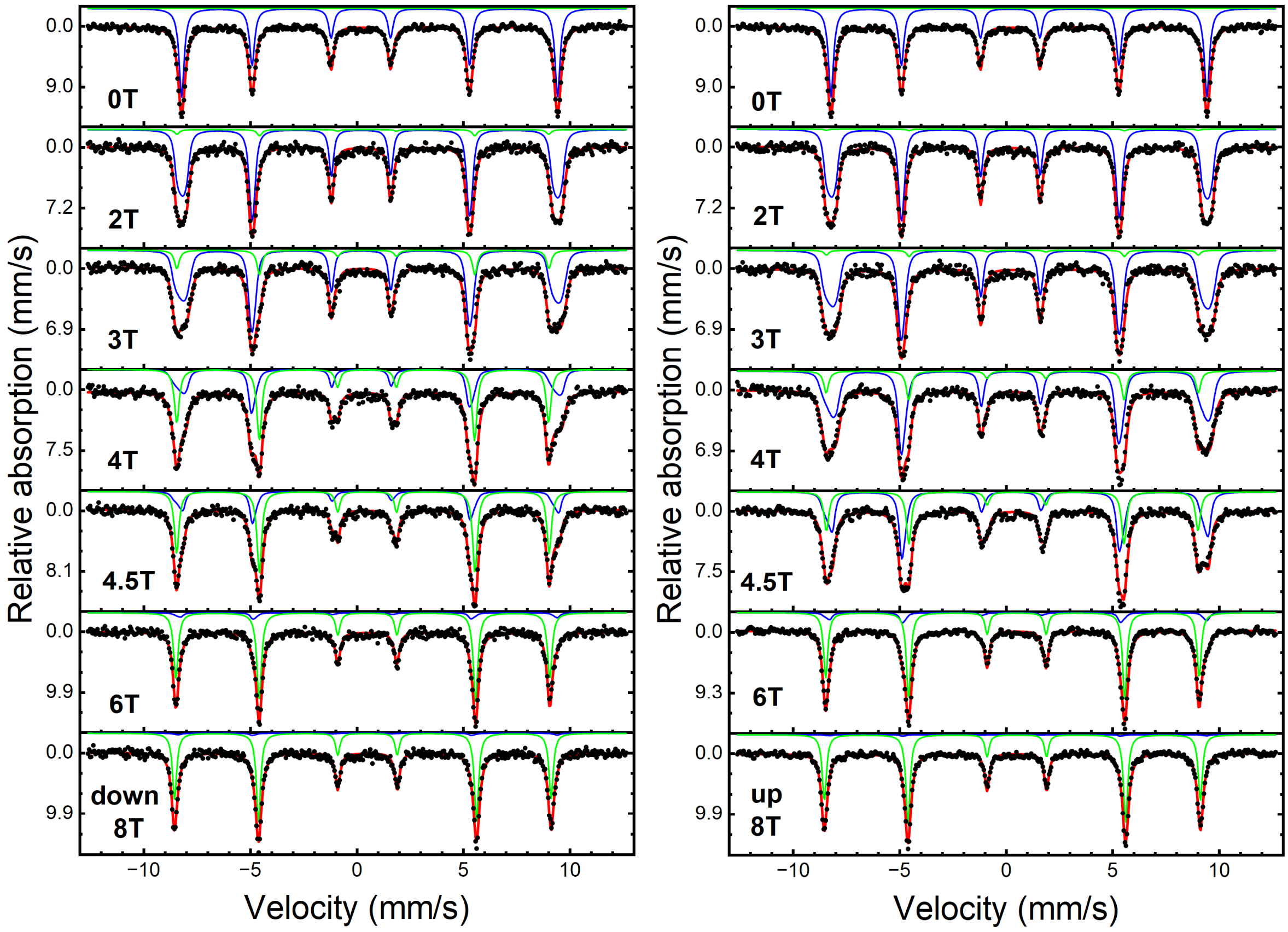

*Figure S7: Mössbauer spectra of sample p = 1.0 recorded at 5 K upon decreasing (left) and increasing (right) magnetic fields between 0 and 8 T.*

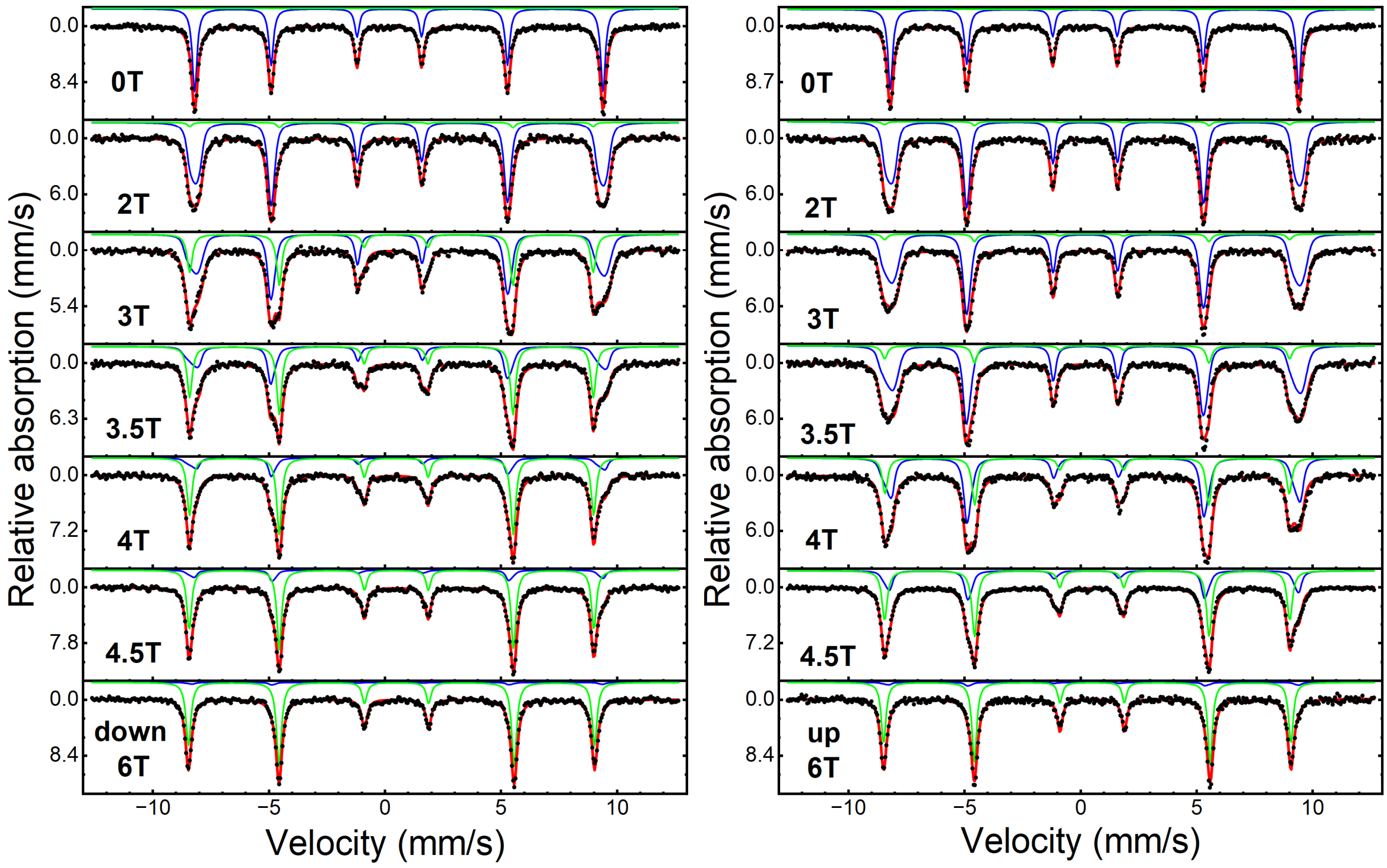

*Figure S8: Mössbauer spectra of sample p = 1.25 recorded at 5 K upon decreasing (left) and increasing (right) magnetic fields between 0 and 8 T.*

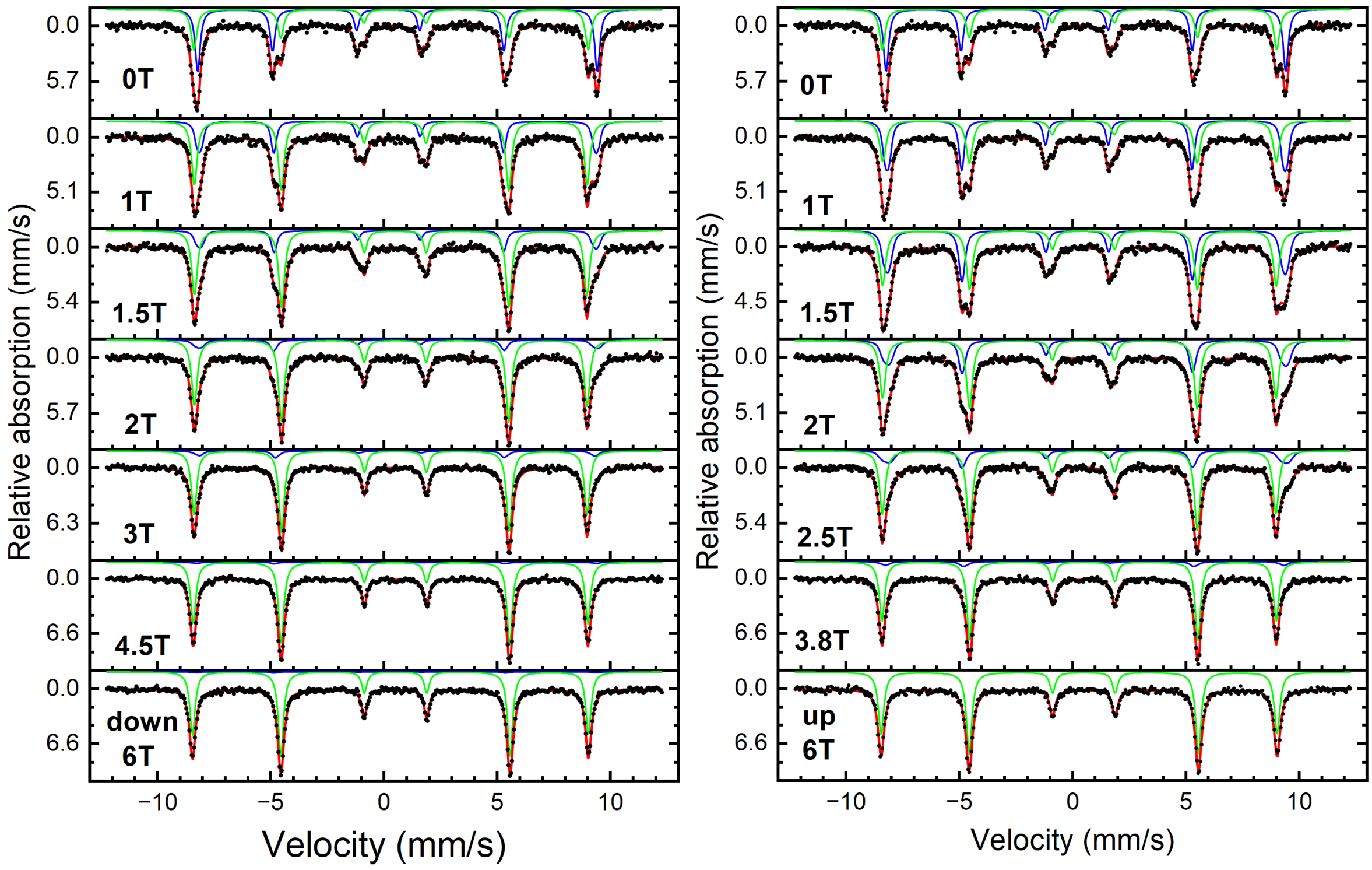

*Figure S9: Mössbauer spectra of sample p = 1.5 recorded at 5 K upon decreasing (left) and increasing (right) magnetic fields between 0 and 6 T.*

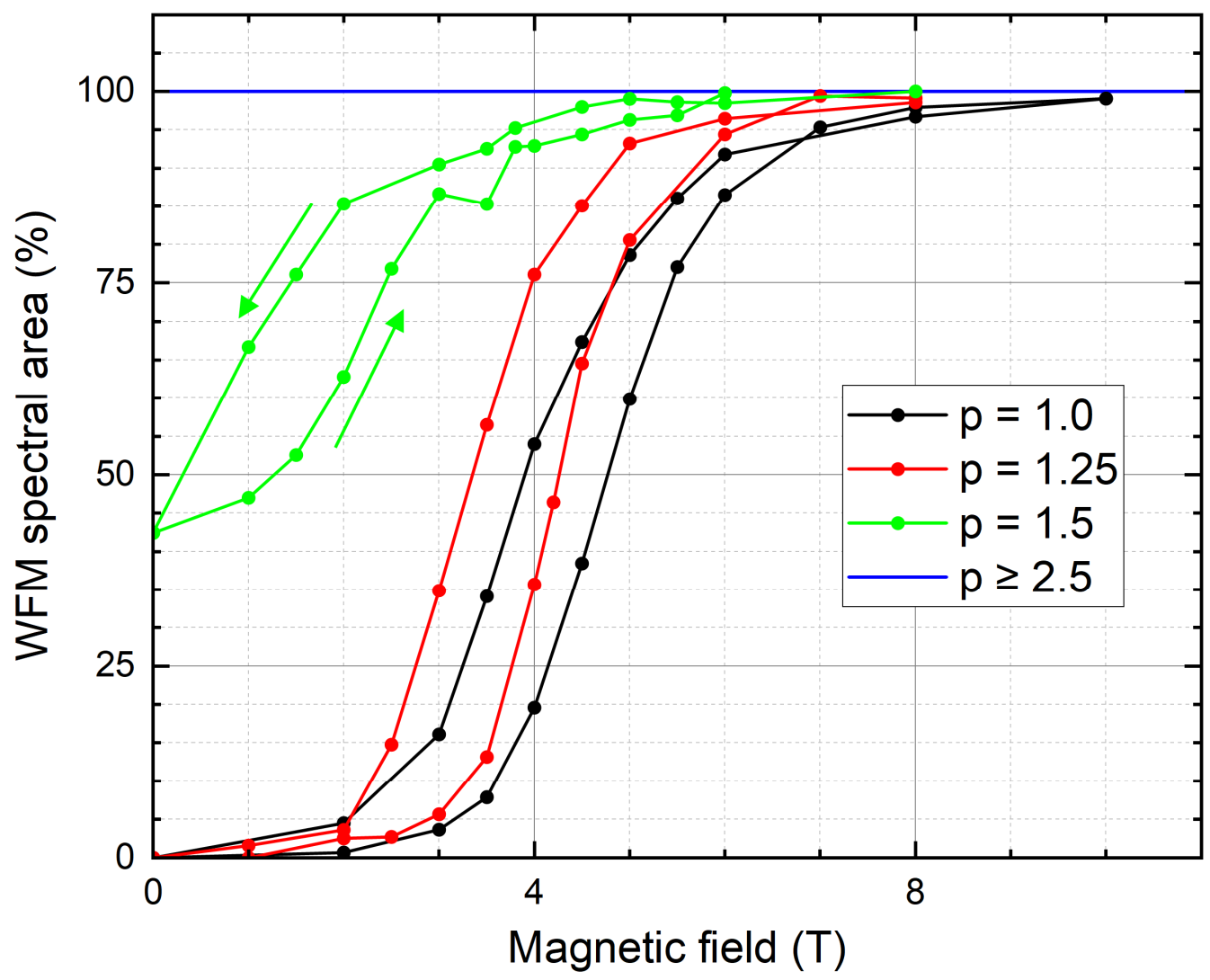

*Figure S10: Weak-ferromagnetic (WFM) spectral area recorded upon rising and decreasing applied magnetic field across the AFM-WFM phase transition via Mössbauer spectroscopy at ca. 5 K for hematite nanospindle aspect ratios of 1.0 (black), 1.25 (red), and 1.5 (green), the blue line represents the stable WFM state for $p \geq 2.5$.*